\documentclass[useAMS,usenatbib]{mn2e}

\usepackage{revsymb}
\usepackage{amsmath}
\usepackage{amsfonts}
\usepackage{amssymb}
\usepackage{graphicx}
\usepackage{lastpage}

\title[A spatial-correlation analysis of the cubic 3-torus topology]
  {A spatial-correlation analysis of the cubic 3-torus topology
   based on the Planck 2013 data}
\author[R. Aurich]
  {R.~Aurich \\
  Institut f\"ur Theoretische Physik, Universit\"at Ulm,\\
  Albert-Einstein-Allee 11,\\ D-89069 Ulm, Germany
}

\begin{document}

\date{}

\pagerange{\pageref{firstpage}--\pageref{LastPage}} \pubyear{2014}

\def\LaTeX{L\kern-.36em\raise.3ex\hbox{a}\kern-.15em
    T\kern-.1667em\lower.7ex\hbox{E}\kern-.125emX}

\newtheorem{theorem}{Theorem}[section]

\def\bfis{\hbox{\scriptsize\rm i}}
\def\bfi{\hbox{\rm i}}
\def\bfj{\hbox{\rm j}}

\newcommand{\apj}{{Astrophys.\ J.}}
\newcommand{\apjs}{{Astrophys.\ J.\ Supp.}}
\newcommand{\apjl}{{Astrophys.\ J.\ Lett.}}
\newcommand{\aj}{{Astron.\ J.}}
\newcommand{\prl}{{Phys.\ Rev.\ Lett.}}
\newcommand{\prd}{{Phys.\ Rev.\ D}}
\newcommand{\mnras}{{Mon.\ Not.\ R.\ Astron.\ Soc.}}
\newcommand{\araa}{{ARA\&A }}
\newcommand{\aap}{{Astron.\ \& Astrophy.}}
\newcommand{\nat}{{Nature }}
\newcommand{\cqg}{{Class.\ Quantum Grav.}}

\setlength{\topmargin}{-1cm}

\label{firstpage}

\maketitle

\begin{abstract}
Spatial correlations of the cubic 3-torus topology are analysed
using the Planck 2013 data.
The spatial-correlation method for detecting multiply connected spaces is
based on the fact that positions on the cosmic microwave background (CMB) sky,
that are separated by large angular distances,
can be spatially much nearer according to a hypothesized topology.
The comparison of the correlations computed with and without the
assumed topology can reveal whether a promising topological candidate is found.
The level of the expected correlations is estimated by CMB simulations of the
cubic 3-torus topology and compared to those obtained from the Planck data.
An interesting 3-torus configuration is discovered
which possesses topological correlations of the magnitude
found in the CMB simulations based on a toroidal universe.
Although the spatial-correlation method has a high false-positive rate,
it is striking that there exists an orientation of a cubic 3-torus cell,
where correlations between points
that are separated by large angular distances,
mimic those of closely separated points.
\end{abstract}



\section{Introduction}

The cosmic microwave background (CMB) radiation is of utmost
importance to constrain the parameters of the cosmological concordance
model, the $\Lambda$CDM model,
because it reveals the state of the cosmos at a very early moment
in its evolution.
But its importance is not only due to this early snap-shot,
since it also contains information of the cosmos on its largest
scales.
Therefore, the CMB radiation might also reveal the topology of the Universe,
that is whether the space is simply connected or multiply connected.
Apart from general considerations, the topic of cosmic topology
\citep{Lachieze-Rey_Luminet_1995,Luminet_Roukema_1999,Levin_2002,%
Reboucas_Gomero_2004,Mota_Reboucas_Tavakol_2010,%
Mota_Reboucas_Tavakol_2011,Fujii_Yoshii_2011}
experiences a boost after the discovery of very low temperature correlations
in the CMB by \cite{Hinshaw_et_al_1996}.
This property is present in all subsequent CMB observations, see
\cite{Copi_Huterer_Schwarz_Starkman_2008,Copi_Huterer_Schwarz_Starkman_2013a}
for a comparison of different data sets.
The low temperature correlations are most obviously seen in
the temperature two-point angular correlation function
\begin{equation}
\label{Eq:C_theta}
C(\vartheta) \; := \; \left< \delta T(\hat n) \delta T(\hat n')\right>
\hspace{10pt} \hbox{with} \hspace{10pt}
\hat n \cdot \hat n' = \cos\vartheta
\hspace{10pt} ,
\end{equation}
where $\vartheta$ is the angular distance between the two directions
$\hat n$ and $\hat n'$ on the surface of last scattering.
The brackets denote averaging over the pixel pairs with
$\hat n \cdot \hat n' = \cos\vartheta$.
However, the central quantity in this paper is not the angular correlation
(\ref{Eq:C_theta}), but the spatial-correlation function $\xi(r)$,
where $r$ is the spatial distance between the two positions
$\hat n$ and $\hat n'$ on the surface of last scattering.
The spatial-correlation function $\xi(r)$ can be used as a tool for
discovering a non-trivial topology of the Universe
as suggested by \cite{Roukema_et_al_2008a}.
The idea is that the distance between two points $q_1$ and $q_2$
is not uniquely defined in a multiply connected space.
On the one hand, one can compute the distance by completely ignoring
the topology which means that one calculates the distance $d(q_1,q_2)$
in the universal cover.
This leads to the definition of the correlation function
\begin{equation}
\label{Eq:corr_triv}
\xi_{\hbox{\scriptsize triv}}(r) :=
\frac{1}{C(0)}
\left< \delta T(q_1) \, \delta T(q_2) \right>
\hspace{3pt}  \hbox{with} \hspace{3pt}
r \; = \; d(q_1,q_2)
\hspace{3pt} ,
\end{equation}
where $C(0)$ normalizes the spatial correlation
$\xi_{\hbox{\scriptsize triv}}(0)=1$.
On the other hand, one can use the shortest possible distance between
$q_1$ and $q_2$ according to the assumed topological structure.
This minimal distance might be computed by
\begin{equation}
\label{Eq:dist_topo_with unit}
d_\Gamma(q_1,q_2) :=
\min_{\gamma\in\Gamma} \, d(q_1,\gamma(q_2))
\hspace{10pt} ,
\end{equation}
where $\Gamma$ denotes the covering group which defines the topology.
However, for reasons that will become clear shortly, 
the topological test uses the slightly modified distance
\begin{equation}
\label{Eq:dist_topo}
d_{\hbox{\scriptsize topo}}(q_1,q_2) :=
\min_{\gamma\in\Gamma'} \, d(q_1,\gamma(q_2))
\hspace{10pt} ,
\end{equation}
where the $\Gamma' := \Gamma \backslash \{\hbox{id}\}$ is the set of
group elements of $\Gamma$ without the identity.
For distances $d(q_1,q_2)$ smaller than the injectivity radius
\citep{Gomero_Reboucas_2003},
one has $d_{\hbox{\scriptsize topo}}(q_1,q_2)>d(q_1,q_2)$,
since the identity is removed.
Removing the identity eliminates the distance $d(q_1,q_2)$ measured
in the trivial topology, i.\,e.\ along the direct path in the universal cover,
from the set of distances over which the minimum is taken in
$d_{\hbox{\scriptsize topo}}(q_1,q_2)$.
Therefore, small distances $d_{\hbox{\scriptsize topo}}(q_1,q_2)$ belong
to widely separated points $q_1$ and $q_2$ on the sky
that are located in different ``copies'' of the fundamental cell.
If $d_{\hbox{\scriptsize topo}}(q_1,q_2)$ is computed for the correct topology,
the distances $d_{\hbox{\scriptsize topo}}(q_1,q_2)$ and $d(q_1,q_2)$
provide two values of the distance between points $q_1$ and $q_2$
measured along different paths.
Now, \cite{Roukema_et_al_2008a} define the topological correlation function
\begin{equation}
\label{Eq:corr_topo}
\xi_{\hbox{\scriptsize topo}}(r) \; := \;
\frac{1}{C(0)}
\left< \delta T(q_1) \, \delta T(q_2) \right>
\hspace{3pt}  \hbox{with} \hspace{3pt}
r \; = \; d_{\hbox{\scriptsize topo}}(q_1,q_2)
\hspace{3pt} ,
\end{equation}
which should reproduce $\xi_{\hbox{\scriptsize triv}}(r)$ under idealized
conditions
although the two functions are computed from different products
$\delta T(q_1) \, \delta T(q_2)$.
At scales $r$ smaller than the injectivity radius,
$\xi_{\hbox{\scriptsize triv}}(r)$ measures the correlations within
the same fundamental cell, while $\xi_{\hbox{\scriptsize topo}}(r)$
measures them between different ``copies'' of the fundamental cell.
Thus the small-scale behaviour $\xi_{\hbox{\scriptsize triv}}(r)$ should
match $\xi_{\hbox{\scriptsize topo}}(r)$ even though the latter is
obtained from widely separated pixels for small values of $r$
as long as $\Gamma$ belongs to the correct topology.
Therefore, the aim of the method is to find a representation of $\Gamma$
such that $\xi_{\hbox{\scriptsize topo}}(r)$ agrees with
$\xi_{\hbox{\scriptsize triv}}(r)$ as far as possible.

It should be mentioned that in the case of a positive signal,
the topology belonging to $\Gamma$ might be confused with a different topology
belonging to $\tilde\Gamma$, if  $\Gamma$ is a subgroup of $\tilde\Gamma$
that is $\Gamma\subset\tilde\Gamma$.
In that case the minimum in equation (\ref{Eq:dist_topo}) would be
taken only over a subset.
In the case of the 3-torus topology, this would mean that the
``true'' fundamental cell would tessellate the one found by the
spatial-correlation method.
Since smaller fundamental cells are easier to detect, this is unlikely
to be the case.

It is worthwhile to note the relation to the matched circle test proposed by
\cite{Cornish_Spergel_Starkman_1998b}.
The positions on a matched circle pair can be mapped onto each other by
a group element $\gamma \in \Gamma'$.
So the distance is $d_{\hbox{\scriptsize topo}}(q_1,q_2)=0$ in this case.
For zero distance $r=0$, $\xi_{\hbox{\scriptsize topo}}(0)$ would thus measure
the correlations between matched circle pairs on the CMB sky.
In this way, $\xi_{\hbox{\scriptsize topo}}(0)$ gives an average of the
correlations over all matched circle pairs.

\cite{Roukema_et_al_2008a} apply their spatial-correlation method to the
Poincar\'e dodecahedral topology which is realized in a space with
positive curvature.
They investigate the three-year data of the Wilkinson Microwave Anisotropy
Probe (WMAP) and find an interesting solution for a special orientation
of the Poincar\'e dodecahedral cell.
The same method is used by \cite{Aurich_2008} with respect to the
cubic 3-torus topology which is the simplest non-trivial topology in
flat space.
The analysis based on the WMAP five-year data leads to a specific
configuration of the cubic 3-torus topology where the spatial correlation
$\xi_{\hbox{\scriptsize topo}}(r)$ is enhanced.
\cite{Roukema_France_Kazimierczak_Buchert_2013} suggest to search for
topologically lensed galaxy pairs in deep red-shift surveys according
to the favoured 3-torus configuration and
simulate the prospects of detecting that topology.
In addition, they show in their figure 1 a variant of the topological
correlation of the favoured orientation using the final WMAP nine-year data
and demonstrate that also the latest WMAP data possess the
spatial-correlation signature in favour of a 3-torus topology.
In this paper the spatial-correlation analysis of the 3-torus topology
is carried out for the Planck 2013 data \citep{Planck_Overview_2013}
in order to see whether the signature persists.

In addition, in section \ref{Sec:Test_for_torus}
the spatial-correlation method is applied to simulated 3-torus CMB maps
so that the correlations obtained from the Planck 2013 data can be compared
to the expected signal.
The analysis of the Planck data is the topic of
section \ref{Sec:Comparison_with_Planck} and
the final section \ref{Sec:Summary} summarizes and discusses the results.

\section{Test of the spatial-correlation method for the 3-torus topology}
\label{Sec:Test_for_torus}

\begin{figure}
\begin{center}
\begin{minipage}{10cm}
\vspace*{-25pt}
\hspace*{-20pt}\includegraphics[width=10.0cm]{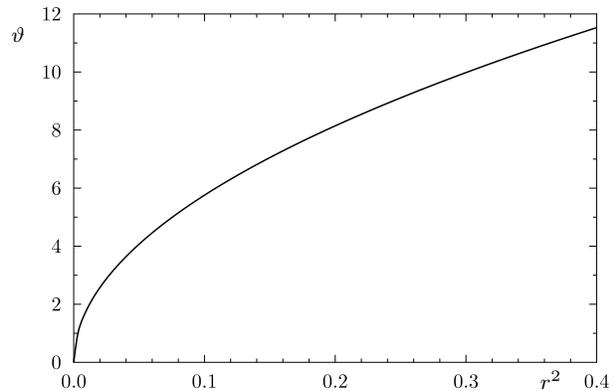}
\put(-65,19){$r^2$}
\put(-265,150){$\vartheta$}
\end{minipage}
\vspace*{-25pt}
\end{center}
\caption{\label{Fig:relation_r_theta}
The projected angle $\vartheta$ onto the surface of last scattering
is plotted in degrees as a function of the squared distance $r^2$.
The diameter $d_{\hbox{\scriptsize SLS}}=6.2972$
of the surface of last scattering is used.
}
\end{figure}

The spatial-correlation method is tested for the cubic 3-torus topology.
To that aim, torus CMB maps are computed for the $\Lambda$CDM model with the
parameters given in Table 10 in \cite{Planck_Overview_2013},
column ``Planck+WP+highL+BAO'' with ``Best fit''.
The cosmological parameters are
$\Omega_{\hbox{\scriptsize bar}}= 0.04825$,
$\Omega_{\hbox{\scriptsize cdm}}= 0.2589$,
$\Omega_\Lambda= 0.6928$,
the reduced Hubble constant $h=0.6777$,
the reionisation optical depth $\tau = 0.0953$,
and the scalar spectral index $n_s=0.9611$.
In the following, the side length $L$ of the torus cell and the distance $r$
are given in units of the Hubble length $L_{\hbox{\scriptsize H}}=c/H_0$.
The distance to the surface of last scattering is
$L_{\hbox{\scriptsize SLS}}= \Delta \eta\, L_{\hbox{\scriptsize H}}$
with $\Delta \eta=\eta_0-\eta_{\hbox{\scriptsize rec}}= 3.1486$
and $\eta$ being the conformal time.
The diameter $d_{\hbox{\scriptsize SLS}}$ of the surface of last scattering is
$d_{\hbox{\scriptsize SLS}}=2\Delta \eta=6.2972$ in these units.
It should be pointed out that this value is by a factor $0.95$ smaller
than the value obtained for a $\Lambda$CDM model favoured by the WMAP data
where $\Delta \eta$ lies around 3.33.
One has to bear this in mind when comparing length-scales from papers based
on the WMAP concordance model.
In the following, the spatial correlations will be analysed as
functions of the squared distance $r^2$, whose relation to the angle
$\vartheta = 2\arcsin(r/d_{\hbox{\scriptsize SLS}})$ projected onto the surface
of last scattering is plotted in figure \ref{Fig:relation_r_theta}.
Most of the analyses consider the interval $r\in[0,r_{\hbox{\scriptsize max}}]$
with $r^2_{\hbox{\scriptsize max}}=0.4$ which translates to a projected angle
of $\vartheta_{\hbox{\scriptsize max}}\simeq 11.5^\circ$.

The computation of $d_{\hbox{\scriptsize topo}}(q_1,q_2)$ requires
the group $\Gamma$ which has infinitely many elements
in the case of the toroidal topology.
The numerical analysis takes only the subset of $\Gamma$ into account,
which is obtained by concatenating
at most four times the generators and their inverses.
A larger subset would only be needed for very small
fundamental cells.

With the above cosmological parameters,
100 CMB realisations of the cubic 3-torus topology with side length $L=4$
are generated
where 61\,556\,892 different eigenmodes are taken into account
which belong to the first 50\,000 eigenvalues.
The spherical expansion of the eigenmodes is carried out up to
$l_{\hbox{\scriptsize max}}=1000$.
In addition, 10 further torus simulations are computed which take only
the usual Sachs-Wolfe contribution into account.

It is a common problem for all strategies, which try to detect a
topological signature in the CMB sky,
that only the usual Sachs-Wolfe contribution,
which is proportional to the  gravitational potential,
would reveal a multiply connected space.
Other contributions to the CMB sky do not possess this direct periodicity
and thus reduce the expected topological signal.
The two most important of them are the Doppler contribution and the
integrated Sachs-Wolfe effect.

\begin{figure}
\begin{center}
\begin{minipage}{10cm}
\vspace*{-25pt}
\hspace*{-20pt}\includegraphics[width=10.0cm]{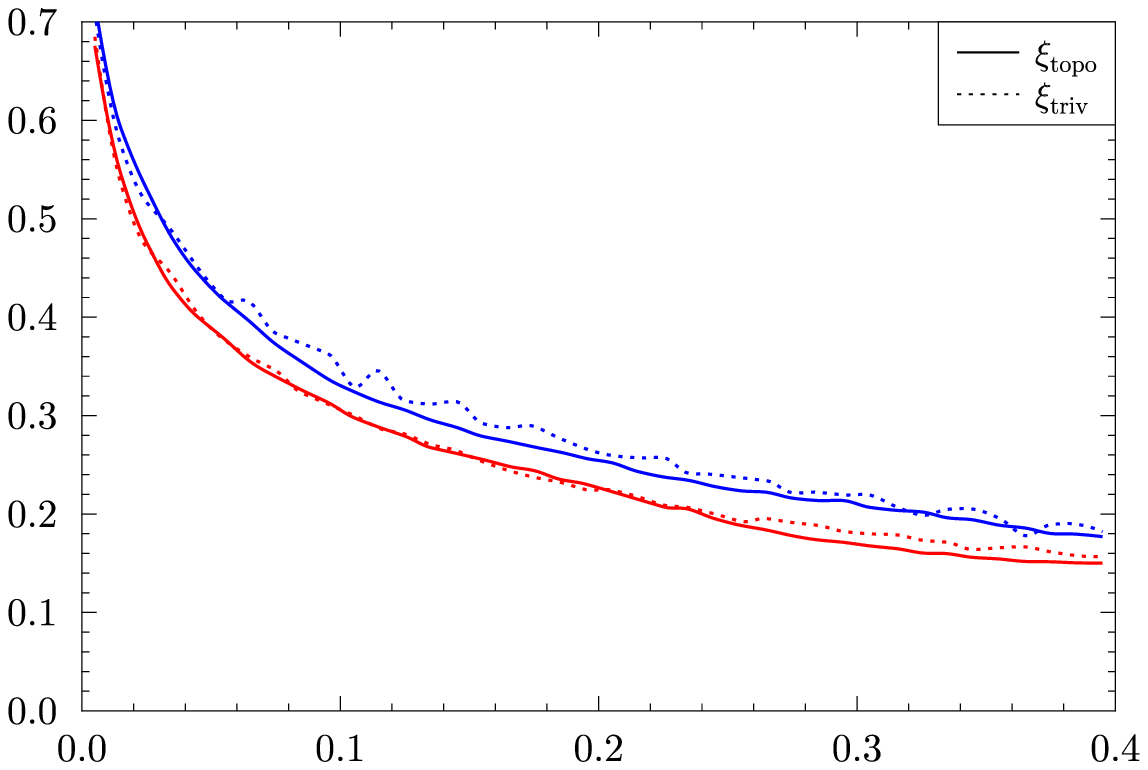}
\put(-210,140){(a)}
\put(-65,19){$r^2$}
\put(-270,150){$\xi$}
\end{minipage}
\begin{minipage}{10cm}
\vspace*{-25pt}
\hspace*{-20pt}\includegraphics[width=10.0cm]{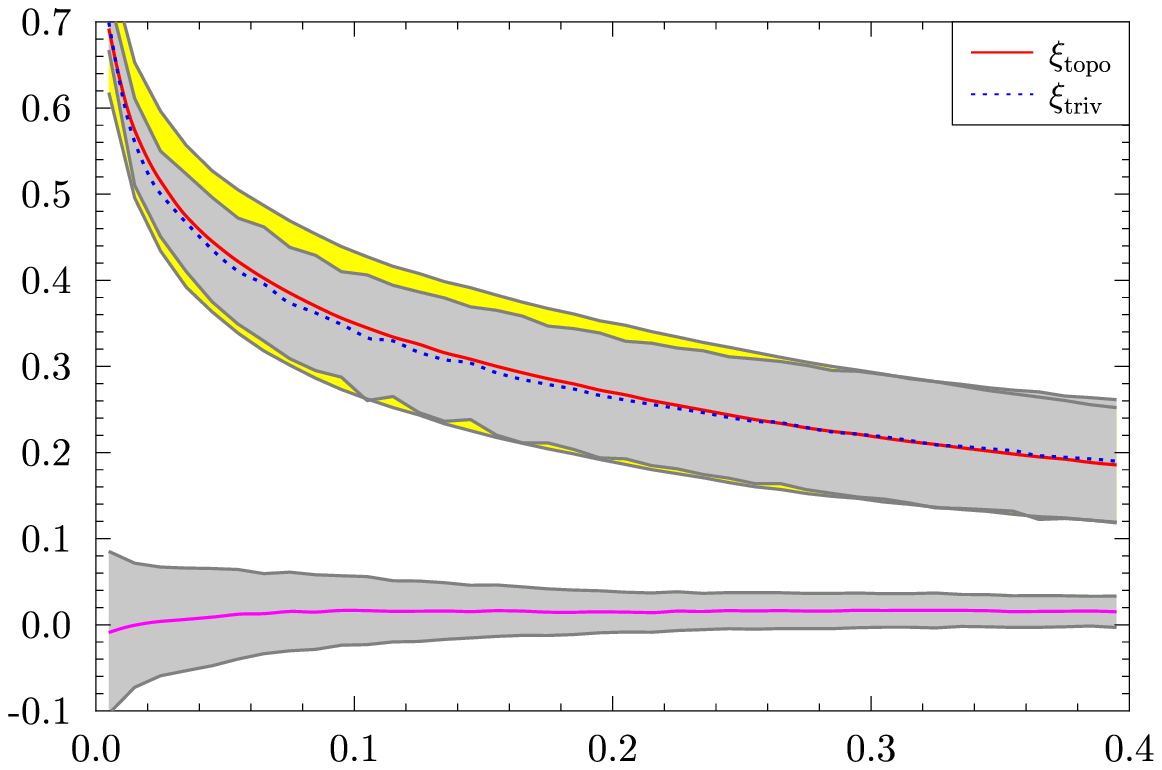}
\put(-210,140){(b)}
\put(-65,19){$r^2$}
\put(-270,150){$\xi$}
\end{minipage}
\vspace*{-25pt}
\end{center}
\caption{\label{Fig:corr_topo_torus_nsw}
The spatial-correlation functions $\xi_{\hbox{\scriptsize triv}}(r)$ and
$\xi_{\hbox{\scriptsize topo}}(r)$ are shown for the 3-torus topology
based on CMB simulations
which take only the usual Sachs-Wolfe contribution into account.
The sky maps have a resolution of $N_{\hbox{\scriptsize side}}=256$, and
a Gaussian smoothing with $\hbox{fwhm}=24'$ is applied.
The Union mask U73 is used.
In panel (a) the result from two simulations is plotted and the good
agreement demonstrates the usability of the method.
Panel (b) shows the distribution of the correlation functions for
10 simulations.
The full curves for $\xi_{\hbox{\scriptsize topo}}(r)$ and the
dotted curves for $\xi_{\hbox{\scriptsize triv}}(r)$ show the mean value
whereas the bands show the $1\sigma$ width of the distributions.
The lower band around zero belongs to correlations
$\xi_{\hbox{\scriptsize topo}}(r)$
which are computed for a wrong configuration of the 3-torus cell.
The upper two overlapping bands belong to $\xi_{\hbox{\scriptsize triv}}(r)$
(the smaller band) and to $\xi_{\hbox{\scriptsize topo}}(r)$ computed for
the orientation used in the simulations.
}
\end{figure}

To demonstrate that difficulty and to test the general procedure,
the spatial-correlation method is at first applied to the 10
CMB simulations that are based only on the usual Sachs-Wolfe contribution.
The correlations are computed on sky maps
which have a HEALPix resolution $N_{\hbox{\scriptsize side}}=256$ and
which are  Gaussian smoothed with a width $\hbox{fwhm}=24'$.
Furthermore, the averaging in (\ref{Eq:corr_triv}) and  (\ref{Eq:corr_topo})
is realized by sampling the products $\delta T(q_1) \, \delta T(q_2)$
in 40 equidistant intervals with respect to the variable $r^2$ up to $r^2=0.4$.
The figure \ref{Fig:corr_topo_torus_nsw}(a) shows for two such simulations
the spatial-correlation functions $\xi_{\hbox{\scriptsize triv}}(r)$
(dotted curve) and $\xi_{\hbox{\scriptsize topo}}(r)$ (full curve).
One observes that $\xi_{\hbox{\scriptsize topo}}(r)$ closely matches
$\xi_{\hbox{\scriptsize triv}}(r)$ although the two are computed from
completely different sets of products $\delta T(q_1) \, \delta T(q_2)$.
The small deviations are due to the pixelisation and smoothing effects.
In figure \ref{Fig:corr_topo_torus_nsw}(b) the mean values
obtained from the 10 functions $\xi_{\hbox{\scriptsize triv}}(r)$ and
$\xi_{\hbox{\scriptsize topo}}(r)$ belonging to the 10 simulations
are plotted as well as the $1\sigma$ band to indicate the width of the
two distributions.
The agreement between $\xi_{\hbox{\scriptsize triv}}(r)$ and
$\xi_{\hbox{\scriptsize topo}}(r)$ is obtained only when the group $\Gamma$,
which is used for the computation of $\xi_{\hbox{\scriptsize topo}}(r)$,
is also the group for which the CMB simulation is computed.
This is demonstrated for another torus configuration with side length $L=4.41$,
which will become important below.
Applying the wrong group with respect to the torus simulations with $L=4.0$
leads to correlations $\xi_{\hbox{\scriptsize topo}}(r)$ around zero,
which are also shown in figure \ref{Fig:corr_topo_torus_nsw}(b).
This illustrates the principle of the spatial-correlation method
under very idealized conditions.

\begin{figure}
\begin{center}
\begin{minipage}{10cm}
\vspace*{-25pt}
\hspace*{-20pt}\includegraphics[width=10.0cm]{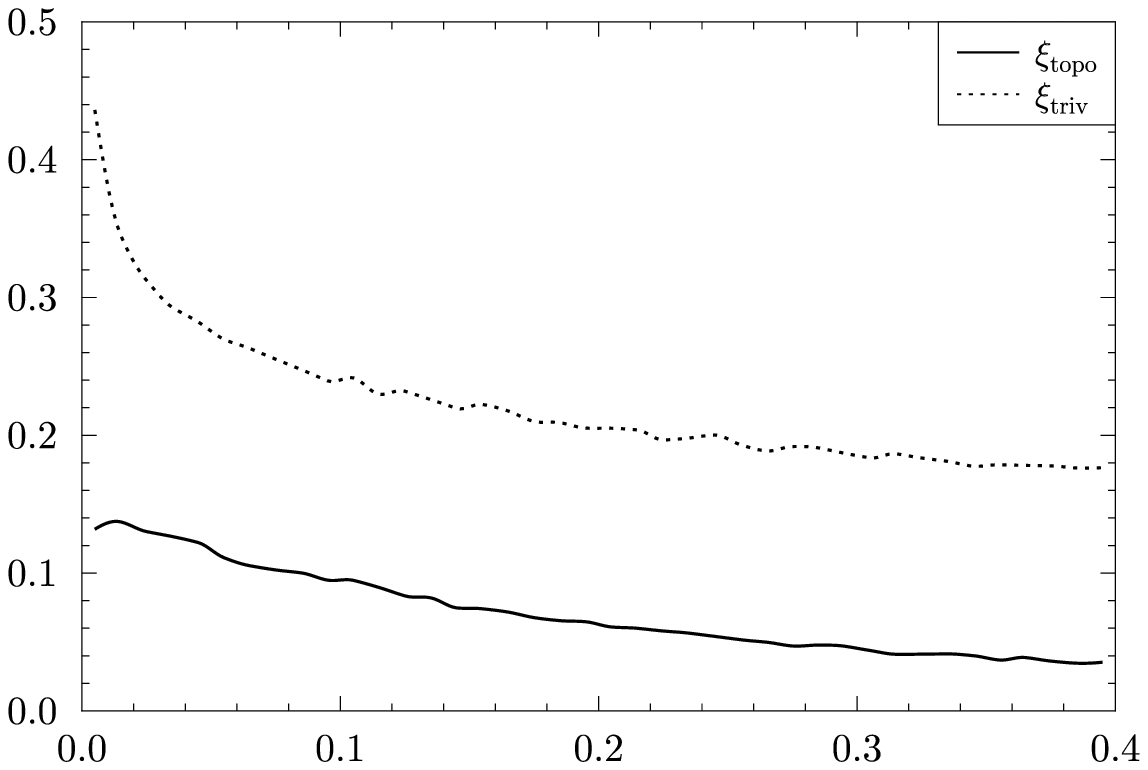}
\put(-210,140){(a)}
\put(-65,19){$r^2$}
\put(-270,150){$\xi$}
\end{minipage}
\begin{minipage}{10cm}
\vspace*{-25pt}
\hspace*{-20pt}\includegraphics[width=10.0cm]{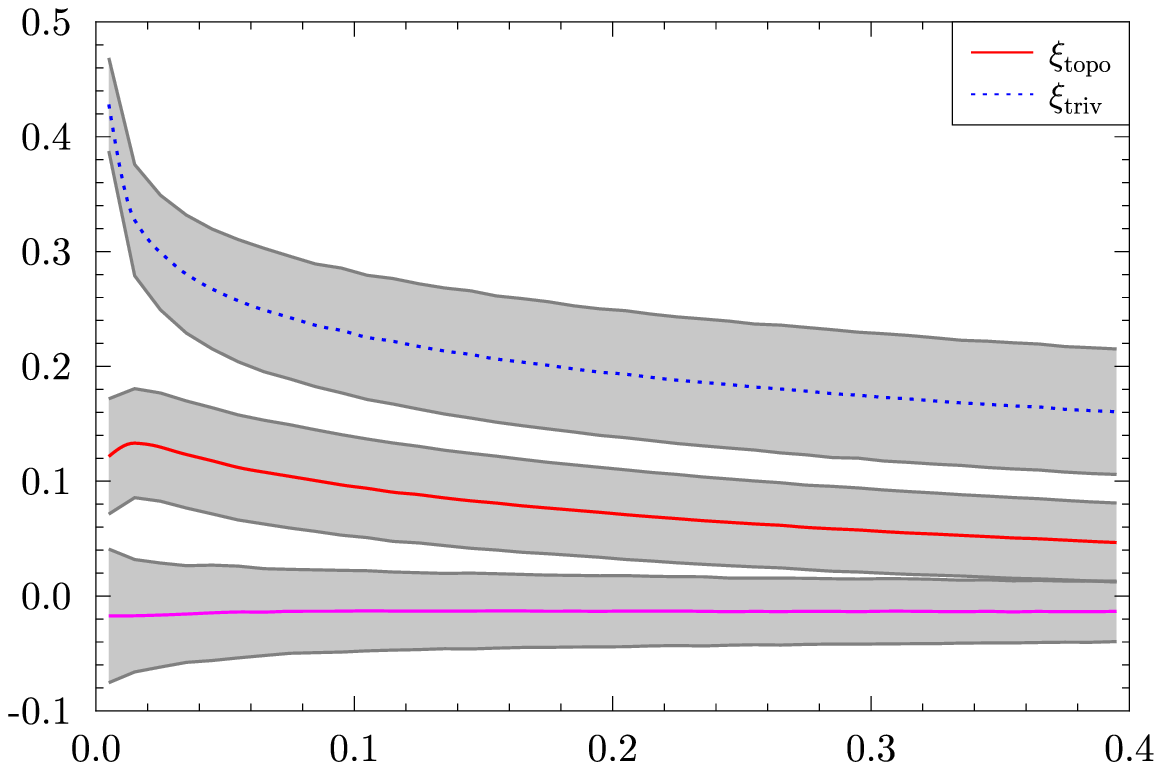}
\put(-210,140){(b)}
\put(-65,19){$r^2$}
\put(-270,150){$\xi$}
\end{minipage}
\vspace*{-25pt}
\end{center}
\caption{\label{Fig:corr_topo_torus}
The spatial-correlation functions $\xi_{\hbox{\scriptsize triv}}(r)$ and
$\xi_{\hbox{\scriptsize topo}}(r)$ are shown for the same 3-torus topology
as in figure \ref{Fig:corr_topo_torus_nsw}.
However, in contrast to figure \ref{Fig:corr_topo_torus_nsw}, the CMB
simulations are now based on the full Boltzmann code taking the deteriorating
effects into account.
The resolution of the maps is again $N_{\hbox{\scriptsize side}}=256$ and
$\hbox{fwhm}=24'$.
The Union mask U73 is applied.
Panel (a) reveals that the match between $\xi_{\hbox{\scriptsize triv}}(r)$
(dotted curve) and $\xi_{\hbox{\scriptsize topo}}(r)$ (full curve) is now
much poorer than in figure \ref{Fig:corr_topo_torus_nsw}(a) where only
the usual Sachs-Wolfe contribution was considered, of course.
Similar to figure \ref{Fig:corr_topo_torus_nsw}(b), the panel (b) displays the
distribution obtained from 100 3-torus simulations.
The $1\sigma$ bands obtained form $\xi_{\hbox{\scriptsize triv}}(r)$ and
from $\xi_{\hbox{\scriptsize topo}}(r)$ using the correct orientation
no longer overlap.
But the correlation $\xi_{\hbox{\scriptsize topo}}(r)$ is nevertheless
significantly larger than those obtained from a wrong 3-torus orientation,
as the lowest band around zero demonstrates.
}
\end{figure}

The CMB radiation is, however, not only described by the usual
Sachs-Wolfe contribution since a variety of physical effects modify it
which destroys the idealized situation.
Thus, 100 torus simulations are computed using a transfer function
which takes the full Boltzmann physics into account.
This includes the Doppler contribution, the integrated Sachs-Wolfe effect,
Silk damping, reionisation, polarisation of photons and neutrinos
assuming a standard thermal history.
In this realistic scenario, the good agreement between
$\xi_{\hbox{\scriptsize triv}}(r)$ and $\xi_{\hbox{\scriptsize topo}}(r)$,
which is observed in figure \ref{Fig:corr_topo_torus_nsw}, is lost.
In figure \ref{Fig:corr_topo_torus}(a) the spatial correlations
$\xi_{\hbox{\scriptsize triv}}(r)$ and $\xi_{\hbox{\scriptsize topo}}(r)$
are plotted for such a realistic simulation.
The correlation $\xi_{\hbox{\scriptsize topo}}(r)$ is computed by
using the same group $\Gamma$ which is also used in the calculation
of the torus sky map.
The figure \ref{Fig:corr_topo_torus}(a) reveals the extend to which the
two correlation functions agree if the correct topology is used
in $\xi_{\hbox{\scriptsize topo}}(r)$.
Furthermore, a comparison between figures \ref{Fig:corr_topo_torus_nsw}(a) and
\ref{Fig:corr_topo_torus}(a) shows
that $\xi_{\hbox{\scriptsize triv}}(r)$ decreases with the
inclusion of the full physics faster than without, 
although in both cases they are normalized by
$\xi_{\hbox{\scriptsize triv}}(0)=1$.
Similar to figure \ref{Fig:corr_topo_torus_nsw}(b),
the figure \ref{Fig:corr_topo_torus}(b) shows the average and the $1\sigma$
bands of the 100 torus simulations but now using the full transfer function.
The three bands are, from top to bottom,
the $1\sigma$ band for $\xi_{\hbox{\scriptsize triv}}(r)$,
the $1\sigma$ band for $\xi_{\hbox{\scriptsize topo}}(r)$ using the
correct configuration of the 3-torus cell, and
the $1\sigma$ band for $\xi_{\hbox{\scriptsize topo}}(r)$ computed
for a wrong configuration.

\begin{figure}
\begin{center}
\begin{minipage}{10cm}
\vspace*{-25pt}
\hspace*{-20pt}\includegraphics[width=10.0cm]{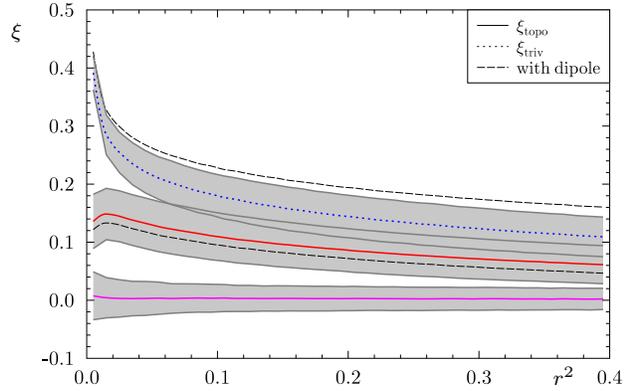}
\put(-65,19){$r^2$}
\put(-270,150){$\xi$}
\end{minipage}
\vspace*{-25pt}
\end{center}
\caption{\label{Fig:corr_topo_torus_no_dipol}
The spatial-correlation functions $\xi_{\hbox{\scriptsize triv}}(r)$ and
$\xi_{\hbox{\scriptsize topo}}(r)$ are plotted for the same 3-torus simulations
as in figure \ref{Fig:corr_topo_torus}(b),
but now the dipole contribution is set to zero.
To guide the eye, the mean of $\xi_{\hbox{\scriptsize triv}}(r)$ and
$\xi_{\hbox{\scriptsize topo}}(r)$ with the inclusion of the dipole are
also shown as dashed curves.
These are the same curves as in figure \ref{Fig:corr_topo_torus}(b).
}
\end{figure}

There is a further complication in the search for a topological signature
in the spatial-correlation function.
In the observed sky maps the dipole contribution is usually removed
because of the difficulty to separate the dynamic dipole due to our motion
with respect to the CMB from the intrinsic dipole contribution
(see e.\,g.\ \cite{Atrio_Barandela_et_al_2014} and references therein).
The 3-torus simulations do contain the intrinsic dipole.
Therefore, in order to compare the Planck data with the simulations,
one has to subtract the dipole in the simulations before computing
the spatial-correlation functions.
The change in the correlation functions is shown in
figure \ref{Fig:corr_topo_torus_no_dipol},
where the spatial-correlation functions are computed from the same
100 simulations as in figure \ref{Fig:corr_topo_torus}(b)
but now without the dipole contribution.
The correlation $\xi_{\hbox{\scriptsize triv}}(r)$ decreases since a
non-vanishing dipole leads to enhanced values in the correlation
(\ref{Eq:corr_triv}) for the considered small values of $r$
belonging to two points $q_1$ and $q_2$ having a small angular separation.
The reverse behaviour occurs for $\xi_{\hbox{\scriptsize topo}}(r)$,
where small values of $r$ belong to large angular separations
where the dipole contribution has most likely a different sign at
the two widely separated points $q_1$ and $q_2$.
This behaviour can be inferred from figure \ref{Fig:corr_topo_torus_no_dipol}
where the mean correlations computed from sky maps with the dipole are also
plotted (dashed curves). 
The $1\sigma$ bands obtained from the 100 simulations without the
dipole contribution will be used below
when the spatial correlations in the Planck data are analysed.
In most cases, these are the $1\sigma$ bands of
figure \ref{Fig:corr_topo_torus_no_dipol}
where the resolution $\hbox{fwhm}=24'$ is used.
When CMB data are analysed in another resolution,
the $1\sigma$ bands are computed from simulations with the corresponding
resolution.

\begin{figure}
\begin{center}
\begin{minipage}{10cm}
\vspace*{-25pt}
\hspace*{-20pt}\includegraphics[width=10.0cm]{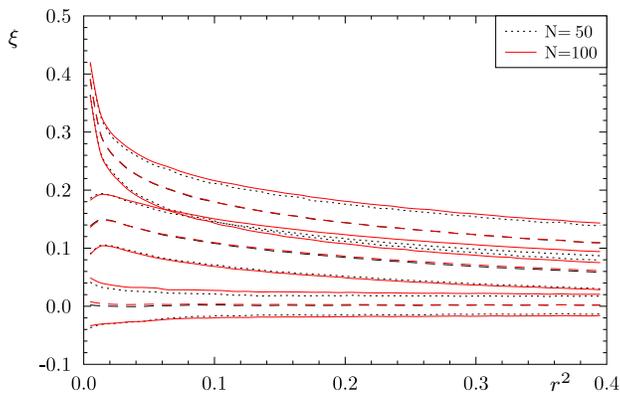}
\put(-65,19){$r^2$}
\put(-270,150){$\xi$}
\end{minipage}
\vspace*{-25pt}
\end{center}
\caption{\label{Fig:corr_topo_torus_50_vs_100}
The mean values and the $1\sigma$ boundaries of the spatial-correlation functions
$\xi_{\hbox{\scriptsize triv}}(r)$ and $\xi_{\hbox{\scriptsize topo}}(r)$ are plotted
for the same 3-torus simulations as in figure \ref{Fig:corr_topo_torus_no_dipol}.
The dotted and full curves are based on the first 50 simulations and all 100
simulations, respectively.
The figure shows that the mean values are almost unchanged while the slight
differences in the $1\sigma$ boundaries reveal the order of the accuracy
obtained from 100 simulations.
}
\end{figure}

It might seem that the number of 100 CMB simulations is too small in order
to extract the mean values and the $1\sigma$ bands belonging to the
spatial-correlation functions of the 3-torus topology.
In order to address this issue,
figure \ref{Fig:corr_topo_torus_50_vs_100} displays these curves
obtained from the first 50 simulations from the set of the 100 simulations
together with the results obtained from all 100 simulations.
The mean values are nearly indistinguishable so that no significant
alteration is expected for an even larger set of simulations.
The $1\sigma$ boundaries show minute differences which indicate
the accuracy that can be achieved from 100 simulations.

It is worthwhile to note that $\xi_{\hbox{\scriptsize topo}}(r)$,
computed for the correct configuration, decreases
at very small distances $r$ as seen in figure \ref{Fig:corr_topo_torus}(a).
This tendency can also be seen for the average and the $1\sigma$ band in
figures \ref{Fig:corr_topo_torus}(b) and \ref{Fig:corr_topo_torus_no_dipol}.
This contrasts to $\xi_{\hbox{\scriptsize triv}}(r)$ which has its maximum
at $r=0$.
Since this behaviour does not occur in figure \ref{Fig:corr_topo_torus_nsw}
dealing with the pure usual Sachs-Wolfe contribution,
it has to be ascribed to the deteriorating contributions contained in
the full transfer function.
Therefore, the degrading effects are less severe for slightly larger
topological distances $r$.
This shows that the spatial-correlation analysis can be a useful
additional tool,
since the matched circle test relies on $\xi_{\hbox{\scriptsize topo}}(r)$
at $r=0$ as discussed in the introduction.

\section{Toroidal correlations in the Planck data}
\label{Sec:Comparison_with_Planck}

\begin{figure}
\begin{center}
\begin{minipage}{10cm}
\vspace*{-25pt}
\hspace*{-20pt}\includegraphics[width=10.0cm]{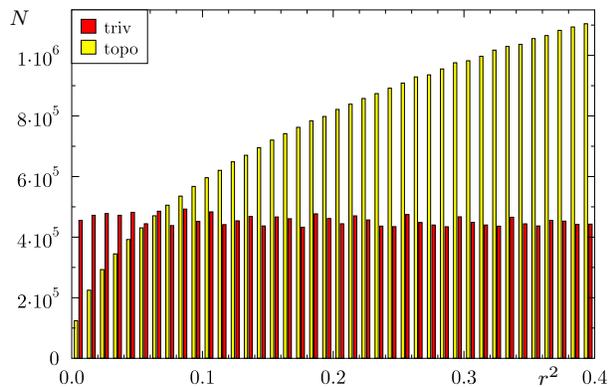}
\put(-65,19){$r^2$}
\put(-265,155){$N$}
\end{minipage}
\vspace*{-25pt}
\end{center}
\caption{\label{Fig:n_corr_nside_256_u73}
The number of pixel pairs are plotted that contribute to the computation of
$\xi_{\hbox{\scriptsize triv}}(i)$ (red bins) and $\xi_{\hbox{\scriptsize topo}}(i)$
(yellow bins), where a 3-torus topology with $L=4$ is tested.
The numbers are obtained from a map with a resolution
$N_{\hbox{\scriptsize side}}=256$ outside the Union mask U73
and every tenth pixel is taken into account.
The interval up to $r^2_{\hbox{\scriptsize max}}=0.4$ is divided into $n=40$ bins.
}
\end{figure}

The correlation analysis of the torus simulations has the advantage
that the underlying topology is known, of course.
In the search for a spatial-correlation signature in the CMB data,
a pseudo-probability estimator is required that measures the
quality of the agreement between $\xi_{\hbox{\scriptsize triv}}(r)$ and
$\xi_{\hbox{\scriptsize topo}}(r)$.
\cite{Roukema_et_al_2008a} define this measure as
\begin{equation}
\label{Eq:P_Roukema}
{\cal P} \; := \; \prod_{i=1}^n \,
\begin{cases}
e^{-\frac{(\xi_{\hbox{\scriptsize topo}}(i)-\xi_{\hbox{\scriptsize triv}}(i))^2}
{2\sigma_i^2}}
& \hbox{ for } \xi_{\hbox{\scriptsize topo}}(i) \leq \xi_{\hbox{\scriptsize triv}}(i) \\
1 + 0.01 \frac{\xi_{\hbox{\scriptsize topo}}(i)-\xi_{\hbox{\scriptsize triv}}(i)}
{\xi_{\hbox{\scriptsize triv}}(i)}
& \hbox{ for }\xi_{\hbox{\scriptsize topo}}(i) \geq \xi_{\hbox{\scriptsize triv}}(i)
\end{cases}
\end{equation}
with
\begin{equation}
\label{Eq:sigma_Roukema}
\sigma_i \; := \; \frac 12 \xi_{\hbox{\scriptsize triv}}(i) \sqrt{\frac{N_n}{N_i}}
\hspace{10pt} .
\end{equation}
Here, the index $i$ runs over the bins for which the values of
$\xi_{\hbox{\scriptsize topo}}(r)$ and $\xi_{\hbox{\scriptsize triv}}(r)$
are sampled.
As described in section \ref{Sec:Test_for_torus}
the $r\in[0,r_{\hbox{\scriptsize max}}]$ interval is divided
into $n=40$ equidistant intervals with respect to the variable $r^2$.
The value $r^2_{\hbox{\scriptsize max}}=0.4$ is generally used 
with the exception of CMB maps having a large smoothing of $\hbox{fwhm}=60'$,
where $r^2_{\hbox{\scriptsize max}}=0.9$ is used, see below.
The number of data points contributing to $\xi_{\hbox{\scriptsize triv}}(i)$
is denoted as $N_i$.
The reason for choosing equidistant intervals with respect to $r^2$ is
that for this binning the values of $N_i$ are of the same order.
This behaviour is revealed by figure \ref{Fig:n_corr_nside_256_u73}
in the case of a map with $N_{\hbox{\scriptsize side}}=256$
where the correlations are computed for every tenth pixel outside
the Union mask U73 for a 3-torus topology with $L=4$.
In addition, figure \ref{Fig:n_corr_nside_256_u73} shows the number
of pixels that contribute to the computation of
$\xi_{\hbox{\scriptsize topo}}(i)$.
Here, the minimum is obtained towards $r^2=0$,
where $\xi_{\hbox{\scriptsize topo}}(i)$ gives an average over all
matched circle pairs.
This demonstrates that a lot of information is used by the
spatial-correlation method beyond the matched circle pairs.

In the search for topological candidates, a modified probability is
used in this paper.
Using $\sigma_i$ as defined in equation (\ref{Eq:sigma_Roukema})
leads to a weighting which disregards bins to which
a small number $N_i$ of data points contribute
or bins with a large value of $\xi_{\hbox{\scriptsize triv}}(i)$.
Thus, this weighting pays less attention to the pronounced peak
in the correlation function at $r=0$.
The peak structure at $r=0$ can be emphasized by modifying the probability
$\ln{\cal P} = -\sum_{i=1}^n w_i\,
\frac{(\xi_{\hbox{\scriptsize topo}}(i)-\xi_{\hbox{\scriptsize triv}}(i))^2}
{2\sigma_i^2}$
with the weight $w_i := \xi_{\hbox{\scriptsize triv}}(i)/
(\frac 1n \sum_{j=1}^n\xi_{\hbox{\scriptsize triv}}(j))$.
This leads to the weighting
\begin{equation}
\label{Eq:sigma_Roukema_modified}
\widehat{\sigma}_i \; := \;
\frac 12 \sqrt{\xi_{\hbox{\scriptsize triv}}(i) \, \frac 1n
\sum_{j=1}^n\xi_{\hbox{\scriptsize triv}}(j)} \; \sqrt{\frac{N_n}{N_i}}
\end{equation}
to be used in equation (\ref{Eq:P_Roukema}).
This discussion shows that there is some arbitrariness in the choice
of the pseudo-probability.
It is clear that this choice determines to some degree
which solution is considered as the optimal one as will be discussed below.

Using the probability (\ref{Eq:P_Roukema}) together with
(\ref{Eq:sigma_Roukema}),
\cite{Roukema_et_al_2008a} and \cite{Aurich_2008} used the
Markov chain Monte Carlo (MCMC) method to search in the WMAP sky maps
for a topological signal in favour of the Poincar\'e dodecahedral topology
and of the cubic 3-torus topology, respectively.
This paper addresses the question whether the cubic 3-torus candidate
found by \cite{Aurich_2008} leads to an enhanced correlation
$\xi_{\hbox{\scriptsize topo}}(r)$ also for the newer CMB data.
The side length of the torus candidate was given as $L\simeq 3.84$
based on the concordance parameters given by WMAP.
As discussed in section \ref{Sec:Test_for_torus}, this corresponds to
$L\simeq 3.66$ using the concordance parameters given by Planck.
The main effect from this shift arises from the lower Hubble constant $H_0$.
The ratio $L/d_{\hbox{\scriptsize SLS}}\simeq 0.58$ with respect to the
diameter of the surface of last scattering is thus the same in both cases.

\begin{figure}
\begin{center}
\begin{minipage}{10cm}
\vspace*{-25pt}
\hspace*{-20pt}\includegraphics[width=10.0cm]{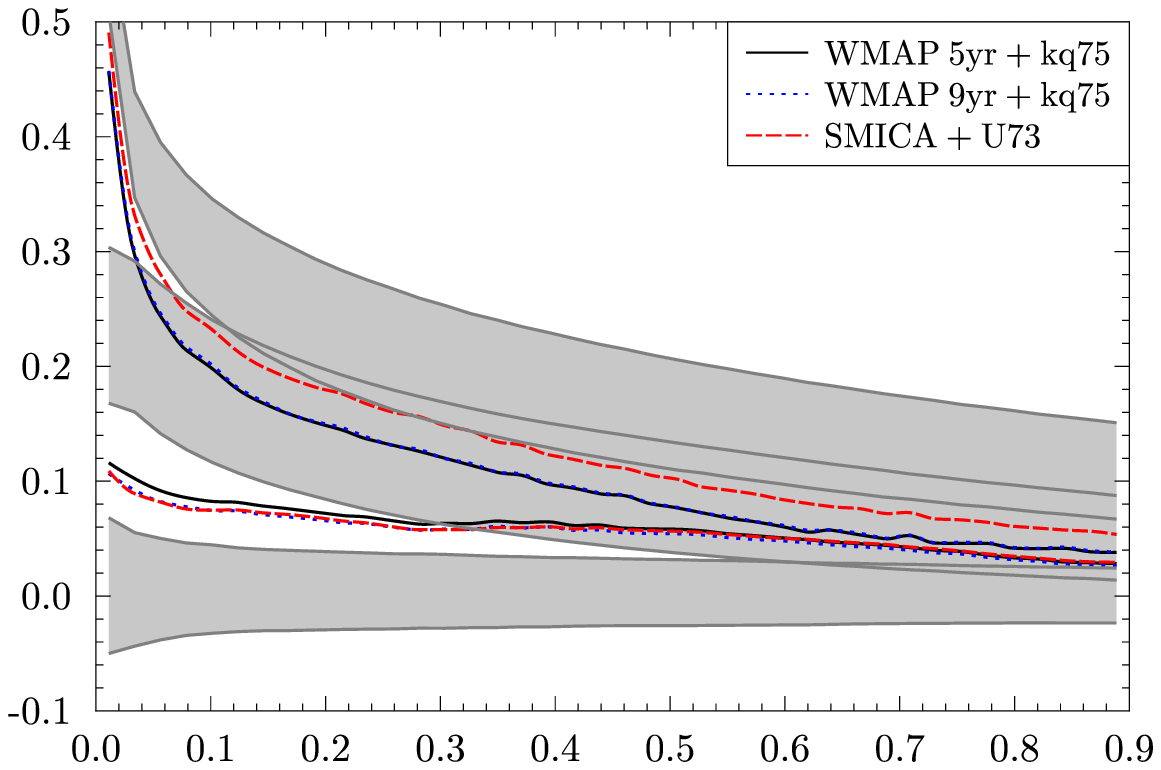}
\put(-210,145){(a)}
\put(-190,145){fwhm = $60'$}
\put(-58,19){$r^2$}
\put(-270,150){$\xi$}
\end{minipage}
\begin{minipage}{10cm}
\vspace*{-25pt}
\hspace*{-20pt}\includegraphics[width=10.0cm]{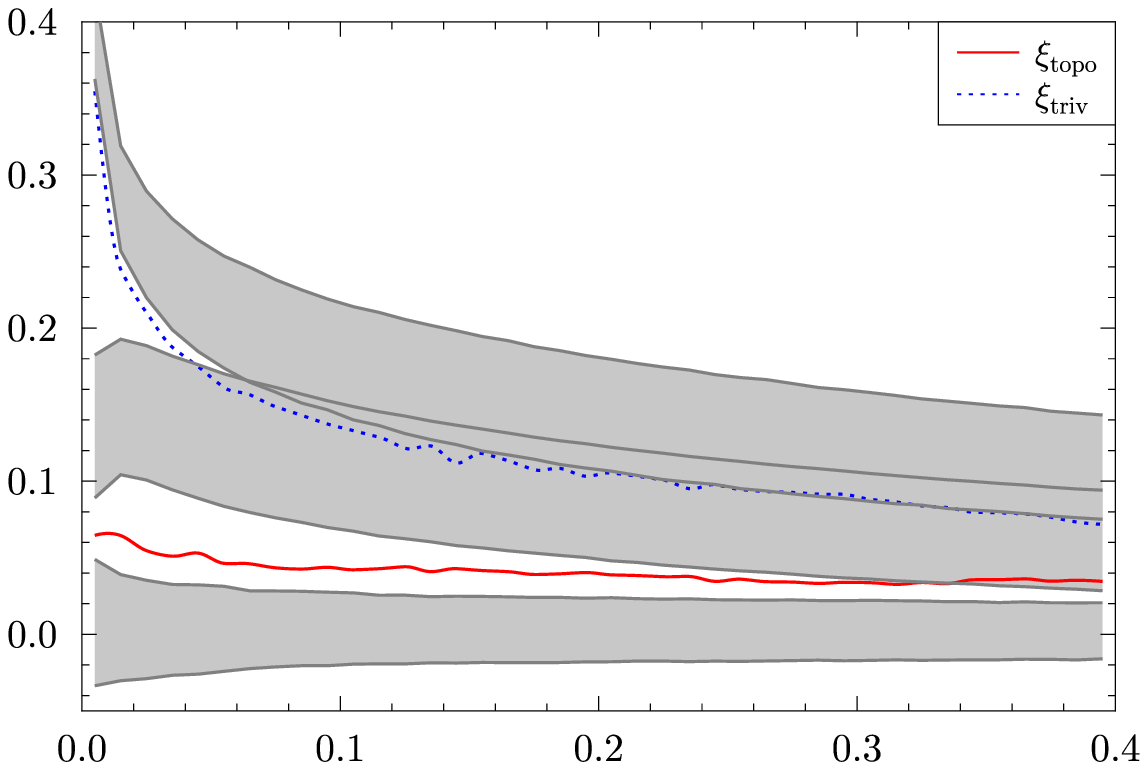}
\put(-210,145){(b)}
\put(-190,145){SMICA with fwhm = $24'$}
\put(-180,132){ + Union mask U73}
\put(-65,19){$r^2$}
\put(-270,150){$\xi$}
\end{minipage}
\vspace*{-25pt}
\end{center}
\caption{\label{Fig:corr_topo_l_3.66}
The correlation $\xi_{\hbox{\scriptsize topo}}(r)$ belonging to the $L=3.66$
torus candidate is shown computed from different data sets.
In panel (a), the correlations obtained from the WMAP five-year data
(full curves)
and the WMAP nine-year data (dotted curves) outside the kq75 mask are plotted
together with correlations of the SMICA map outside the Union mask U73.
The SMICA map is smoothed to the ILC resolution of $60'$.
A pixelisation of $N_{\hbox{\scriptsize side}}=256$ is used in all cases.
The panel (b) displays the correlations for the higher resolution of
$\hbox{fwhm}=24'$ that is possible in the case of the SMICA map.
Both figures show the $1\sigma$ bands computed from the 3-torus simulations
with a resolution of $\hbox{fwhm}=60'$ in panel (a) and
with a resolution of $\hbox{fwhm}=24'$ in panel (b).
}
\end{figure}

In figure \ref{Fig:corr_topo_l_3.66}(a) the spatial correlations
$\xi_{\hbox{\scriptsize triv}}(r)$ and $\xi_{\hbox{\scriptsize topo}}(r)$
are plotted for the cubic 3-torus candidate with $L=3.66$
and the orientation given by \cite{Aurich_2008} based on the
WMAP five-year data \citep{Gold_et_al_2008}.
The figure \ref{Fig:corr_topo_l_3.66}(a) reveals that the WMAP five-year
and the WMAP nine-year data \citep{Bennett_et_al_2012} lead to almost
the same correlations.
Here, the ILC maps outside the kq75 mask are used which have a resolution
of $\hbox{fwhm}=60'$.
In order to compare these correlations with the Planck 2013 data,
the SMICA map is smoothed to $\hbox{fwhm}=60'$, and the correlations
are computed outside the Union U73 mask \citep{Planck_Overview_2013}.
The topological correlation $\xi_{\hbox{\scriptsize topo}}(r)$ is almost
indistinguishable from the WMAP nine-year result,
but the correlation $\xi_{\hbox{\scriptsize triv}}(r)$ lies above the
corresponding WMAP curves.
Note, that $\xi_{\hbox{\scriptsize triv}}(r)$ is independent of any
assumed topology, of course, and that this shift is, therefore,
due to fine structures in the CMB maps.
This effect also occurs in the angular correlation function $C(\vartheta)$,
equation (\ref{Eq:C_theta}), which has different amplitudes
at $\vartheta=0^\circ$ for the WMAP and Planck data.
Figure \ref{Fig:corr_topo_l_3.66}(a) also shows the $1\sigma$ bands obtained
from the 3-torus simulations as described in section \ref{Sec:Test_for_torus}.
It is seen that $\xi_{\hbox{\scriptsize topo}}(r)$ lies for $r^2 \gtrsim 0.35$
within the corresponding $1\sigma$ band, but drops below it for
$r^2 \lesssim 0.35$.
It is, however, above the $1\sigma$ band computed for the wrong
torus configuration and, thus, remains an interesting candidate.

The correlation $\xi_{\hbox{\scriptsize triv}}(r)$ lies below the
$1\sigma$ band of $\xi_{\hbox{\scriptsize triv}}(r)$ obtained
from the torus simulations,
but the deviation from the lower boundary of the $1\sigma$ band is much smaller
for the Planck curve than for the WMAP curves,
see figure \ref{Fig:corr_topo_l_3.66}(a).
This lack of correlations at small distances should not be confused
with the famous lack of correlations at large angles
$\vartheta\gtrsim 60^\circ$ in $C(\vartheta)$, equation (\ref{Eq:C_theta}).
The lack of correlations at small distances in $\xi_{\hbox{\scriptsize triv}}(r)$
is also present in $C(\vartheta)$ at small angles.
Non-trivial topological models are not only able to explain the lack of
correlations at large angles, but also at small angles as shown
by \cite{Aurich_Lustig_Steiner_2004c} for the Poincar\'e dodecahedron and
by \cite{Aurich_Lustig_Steiner_Then_2004b} for the Picard topology.
The lack of correlations below $\vartheta<30^\circ$ is investigated
by \cite{Kim_Naselsky_2011} using the WMAP seven-year data and is found
to be statistically significant.
The analysis of the \cite{Planck_Isotropy_2013} points to a less severe
deviation which is consistent with figure \ref{Fig:corr_topo_l_3.66}(a).
Since this anomaly or deviation does not concern
$\xi_{\hbox{\scriptsize topo}}(r)$,
it is considered hereafter as an observational fact.

Since the SMICA map possesses a resolution of $\hbox{fwhm}=5'$,
it is interesting to see how a better resolution influences the correlations.
In the following figures concerning the spatial-correlation functions,
the distance $r$ is restricted to $r^2\leq 0.4$ in order to emphasize
the peak structure.
For this smaller interval, figure \ref{Fig:corr_topo_l_3.66}(b) displays
the correlations computed from the SMICA map where a Gaussian smoothing
with $\hbox{fwhm}=24'$ is applied and which is downgraded to
$N_{\hbox{\scriptsize side}}=256$.
Furthermore, only pixels outside the Union mask U73 are taken into account.
The correlation $\xi_{\hbox{\scriptsize topo}}(r)$ lies for $r^2 \lesssim 0.3$
below the $1\sigma$ band computed for the correct torus configuration,
but well above the $1\sigma$ band belonging to the wrong configuration.
The correlation $\xi_{\hbox{\scriptsize triv}}(r)$ lies slightly below the
corresponding $1\sigma$ band,
as it is the case for the lower resolution shown in panel (a).
So the conclusion is that this 3-torus configuration is interesting
because it leads to an enhanced topological correlation
$\xi_{\hbox{\scriptsize topo}}(r)$.

\begin{figure}
\begin{center}
\begin{minipage}{10cm}
\vspace*{-25pt}
\hspace*{-20pt}\includegraphics[width=10.0cm]{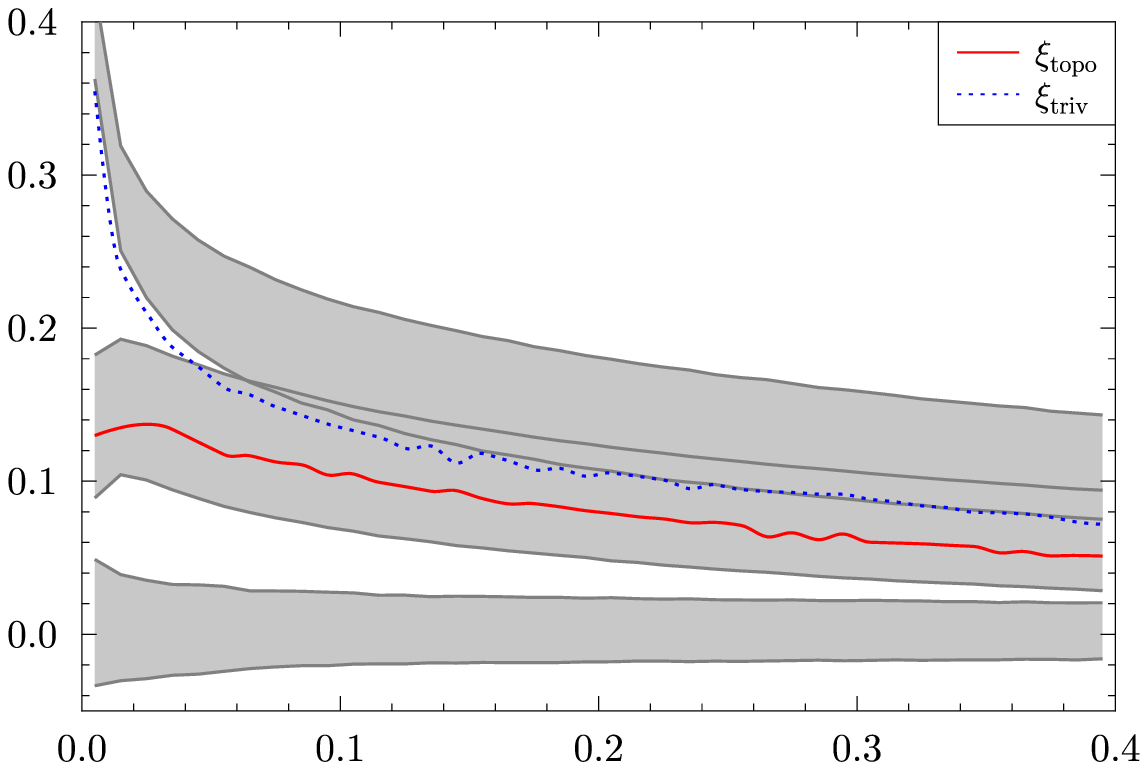}
\put(-210,145){(a)}
\put(-190,145){SMICA with fwhm = $24'$}
\put(-180,132){ + Union mask U73}
\put(-65,19){$r^2$}
\put(-270,150){$\xi$}
\end{minipage}
\begin{minipage}{10cm}
\vspace*{-25pt}
\hspace*{-20pt}\includegraphics[width=10.0cm]{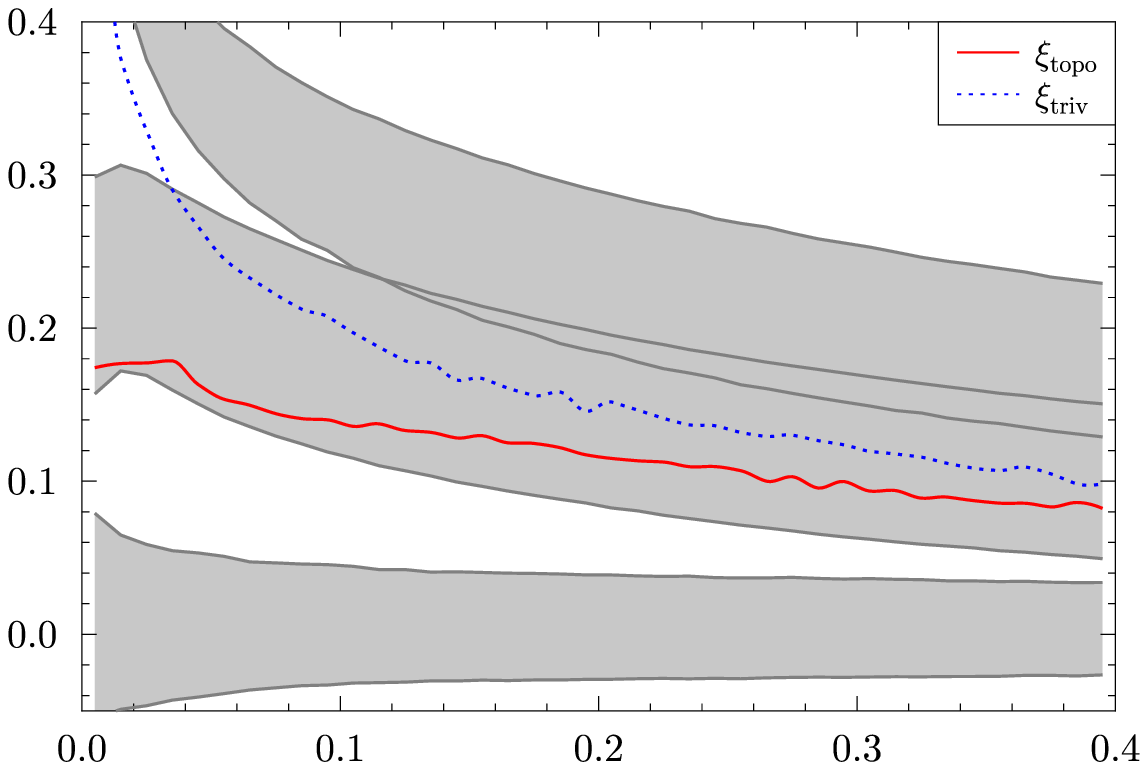}
\put(-192,147){(b)}
\put(-178,147){WMAP 9yr + kq75 mask}
\put(-65,19){$r^2$}
\put(-270,150){$\xi$}
\end{minipage}
\vspace*{-25pt}
\end{center}
\caption{\label{Fig:corr_topo_planck_wmap}
The correlation $\xi_{\hbox{\scriptsize topo}}(r)$ is computed for
the 3-torus candidate with side length $L=4.41$ and the orientation
given in the text.
The spatial correlations obtained from the SMICA map using the Union mask U73
are plotted in panel (a)
and from the WMAP 9yr ILC map using the kq75 mask in panel (b).
The resolution of the maps is again $N_{\hbox{\scriptsize side}}=256$, and
the Planck map is smoothed with $\hbox{fwhm}=24'$.
No smoothing is applied to the ILC map because of its lower resolution
of $60'$.
The $1\sigma$ bands of figure \ref{Fig:corr_topo_torus_no_dipol} are shown
in panel (a).
In panel (b), the $1\sigma$ bands for the lower resolution are plotted.
It is striking to observe in panel (a) that $\xi_{\hbox{\scriptsize topo}}(r)$
obtained from the Planck data almost agrees with the mean topological
correlation derived from the 3-torus simulations for a correct identification.
On the other hand, $\xi_{\hbox{\scriptsize triv}}(r)$ lies below the
corresponding $1\sigma$ band of the simulations.
However, this implies that the match between $\xi_{\hbox{\scriptsize topo}}(r)$
and $\xi_{\hbox{\scriptsize triv}}(r)$ is even better than in many simulations.
The good agreement between $\xi_{\hbox{\scriptsize topo}}(r)$ and
$\xi_{\hbox{\scriptsize triv}}(r)$ is also present in the WMAP 9yr data
as panel (b) reveals using a resolution of $\hbox{fwhm}=60'$.
}
\end{figure}

The SMICA map with the resolution $\hbox{fwhm}=24'$ and
$N_{\hbox{\scriptsize side}}=256$ is also
used to search for further interesting configurations
by applying the MCMC method.
Only pixels outside the Union mask U73 are taken into account.
The parameter space consists of the side length $L$ and the three Euler angles
defining the orientation of the cubic 3-torus cell.
A random point in this four-dimensional parameter space is selected
as the starting point for the MCMC algorithm
which generates a sequence of points with the aim to find more likely
configurations according to the pseudo-probability (\ref{Eq:P_Roukema}).
This method works fine when the region with the enhanced probability
is not too localized, since otherwise the Markov chain can miss
the interesting domain.
Thus, an alternative program without using the MCMC algorithm is carried out
which does a simple random search.
A large set of $N=2\,000\,000$ random points is generated from the parameter
space and the corresponding pseudo-probabilities (\ref{Eq:P_Roukema})
using $\widehat{\sigma}_i$, equation (\ref{Eq:sigma_Roukema_modified}),
are computed in order to find interesting configurations.
The distance interval is again chosen as $r^2 \in [0, 0.4]$.
The parameter space of the random search is confined to $L\in[3.5,5.5]$,
but the three Euler angles are generated without restrictions.
This random search discovers a cubic 3-torus configuration with a
much better agreement with the expected topological signature.
After the random search has found a new more interesting 3-torus candidate,
a MCMC chain is generated, which uses as the starting point the parameters
of this new candidate.
This MCMC chain of length $100\,000$ explores the parameter space around the
candidate and reveals the generators of the group $\Gamma$ for a sample of
3-torus configurations with large correlations $\xi_{\hbox{\scriptsize topo}}(r)$.

\begin{figure}
\begin{center}
\begin{minipage}{10cm}
\vspace*{-25pt}
\hspace*{-20pt}\includegraphics[width=10.0cm]{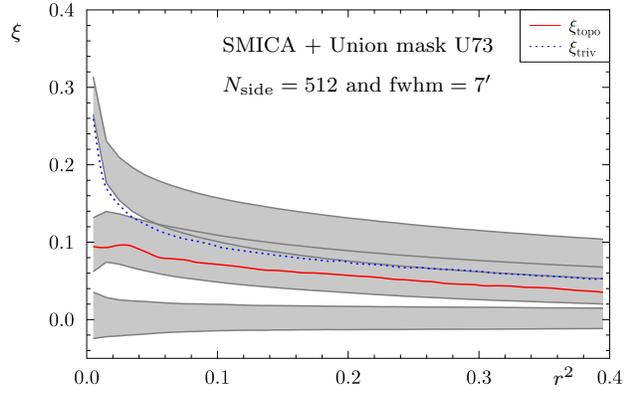}
\put(-190,145){SMICA + Union mask U73}
\put(-190,130){$N_{\hbox{\scriptsize side}}=512$ and $\hbox{fwhm}=7'$}
\put(-65,19){$r^2$}
\put(-270,150){$\xi$}
\end{minipage}
\vspace*{-25pt}
\end{center}
\caption{\label{Fig:corr_topo_planck_512_05}
As in figure \ref{Fig:corr_topo_planck_wmap}(a)
the spatial-correlation functions $\xi_{\hbox{\scriptsize triv}}(r)$ and
$\xi_{\hbox{\scriptsize topo}}(r)$ obtained from
the SMICA map using the Union mask U73 are plotted,
but now the resolutions of all maps are $N_{\hbox{\scriptsize side}}=512$ and
$\hbox{fwhm}=7'$.
The result is very similar to that shown in figure
\ref{Fig:corr_topo_planck_wmap}(a), only the amplitudes are reduced overall.
}
\end{figure}

The new candidate has a side-length $L\simeq 4.41$,
i.\,e.\ $L/d_{\hbox{\scriptsize SLS}}\simeq 0.70$, and the spatial correlations
are shown in figures \ref{Fig:corr_topo_planck_wmap} and
\ref{Fig:corr_topo_planck_512_05}.
Figure \ref{Fig:corr_topo_planck_wmap}(a) displays the correlations
computed from the SMICA map with the resolution $\hbox{fwhm}=24'$ and
$N_{\hbox{\scriptsize side}}=256$ outside the Union mask U73.
The correlation $\xi_{\hbox{\scriptsize triv}}(r)$ is the same as in
figure \ref{Fig:corr_topo_l_3.66}(b) since both figures refer to the
same CMB map.
The topological correlation $\xi_{\hbox{\scriptsize topo}}(r)$ of the
$L=4.41$ candidate is very close to the mean of the correlations
obtained from the torus simulations when the correct orientation is used.
The comparison of figure \ref{Fig:corr_topo_l_3.66}(b) with
\ref{Fig:corr_topo_planck_wmap}(a) reveals the improvement with respect to
the topological signal of the $L=4.41$ candidate.
It is striking to see the agreement between $\xi_{\hbox{\scriptsize topo}}(r)$
and $\xi_{\hbox{\scriptsize triv}}(r)$ in
figure \ref{Fig:corr_topo_planck_wmap}(a).

To answer the question whether this candidate is also distinguished in
the WMAP data,
figure \ref{Fig:corr_topo_planck_wmap}(b) shows the correlations computed
from the WMAP nine-year ILC map outside the kq75 mask.
The ILC map has a resolution of $\hbox{fwhm}=60'$ and the corresponding
$1\sigma$ bands are shown.
As in the case of the Planck data, the agreement between
$\xi_{\hbox{\scriptsize triv}}(r)$ and $\xi_{\hbox{\scriptsize topo}}(r)$
is quite well.
This candidate thus might have been discovered by \cite{Aurich_2008},
but the Markov chains overlooked this interesting domain in the
four-dimensional parameter space and
thus failed to notice this torus configuration.
This demonstrates that searches for topological spaces with more than four
parameters would have to be treated with even more care.

The SMICA map possesses a resolution of $\hbox{fwhm}=5'$.
This map is Gaussian smoothed with $\hbox{fwhm}=5'$ leading to a
map with an effective resolution of $\hbox{fwhm}=7'$.
This map is downgraded to $N_{\hbox{\scriptsize side}}=512$ and the
corresponding spatial correlations are plotted in
figure \ref{Fig:corr_topo_planck_512_05} together with the
appropriate $1\sigma$ bands.
It is seen that the topological correlation
$\xi_{\hbox{\scriptsize topo}}(r)$ is close to
$\xi_{\hbox{\scriptsize triv}}(r)$ also in this higher resolution.

The large values in the correlation $\xi_{\hbox{\scriptsize topo}}(r)$
might be accidentally generated by the structure of the mask for the
special torus configuration with $L\simeq 4.41$.
To provide evidence that this is unlikely, the 3-torus simulations
computed for $L=4.00$ are used
to compute the correlations for the configuration of the $L=4.41$ candidate.
Since this is not the configuration used in the CMB simulations,
one obtains the $1\sigma$ band for the wrong configuration shown in the
previous figures, i.\,e.\ the $1\sigma$ the band around zero.
This disproves the possibility of a spurious correlation somehow
generated by the structure of the mask.

\begin{figure}
\begin{center}
\begin{minipage}{8.6cm}
\includegraphics[width=8.6cm]{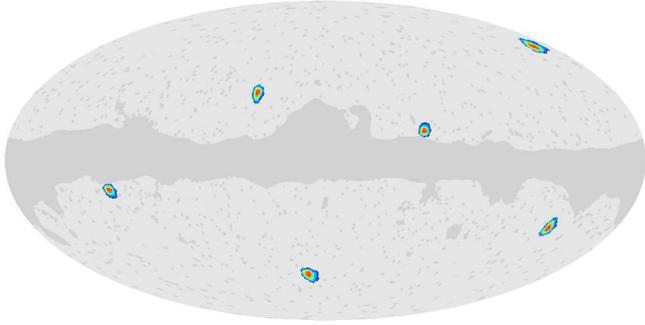}
\end{minipage}
\vspace*{-5pt}
\end{center}
\caption{\label{Fig:mollweide_axis}
The orientation of the fundamental symmetry axes of the cubic 3-torus candidate
around $L=4.41$ are shown using the Mollweide projection.
Galactic coordinates are used and the Galactic centre lies in the
centre of the plot.
The Union mask U73 is also indicated by the shaded region.
}
\end{figure}

\begin{table}
\centering
\begin{tabular}{|c|c|c|c|c|c|c|}
\hline
$l$ &  13$^\circ$ &  42$^\circ$ & 123$^\circ$ & 193$^\circ$ & 222$^\circ$ & 303$^\circ$ \\
\hline
$b$ & -55$^\circ$ &  31$^\circ$ & -14$^\circ$ &  55$^\circ$ & -31$^\circ$ &  14$^\circ$ \\
\hline
\end{tabular}
\caption{\label{Tab:Torus_axes}
The positions of the fundamental symmetry axes of the cubic 3-torus candidate
are listed in Galactic coordinates $(l,b)$.
The values are rounded to one degree.
}
\end{table}

The MCMC chain, which explores the parameter space
around the $L=4.41$ candidate, can be used to reveal the orientation
of the cubic 3-torus cell.
The three generators of $\Gamma$ define three axes through
our observer position, which intersect the surface of last scattering at
six positions.
The positions of these fundamental directions are shown in
figure \ref{Fig:mollweide_axis},
where the full sky is mapped by the Mollweide projection
using Galactic coordinates.
The intensities are plotted according to ${\cal P}$ using $\widehat{\sigma}_i$
as defined in equation (\ref{Eq:sigma_Roukema_modified}),
which is also used for the computation of the MCMC chain.
The positions of the six fundamental directions are given
in table \ref{Tab:Torus_axes}.

\begin{figure}
\begin{center}
\begin{minipage}{8.6cm}
\vspace*{-25pt}
\includegraphics[width=8.6cm]{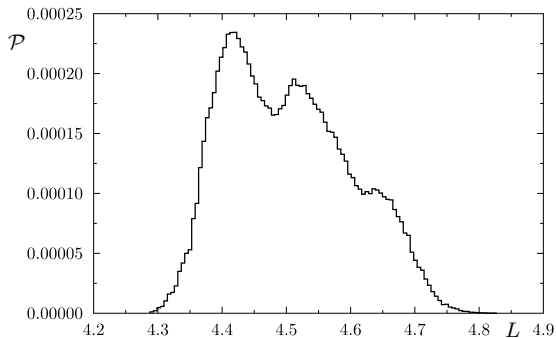}
\put(-52,16){$L$}
\put(-240,125){${\cal P}$}
\end{minipage}
\vspace*{-15pt}
\end{center}
\caption{\label{Fig:P_as_function_of_L}
The largest values of ${\cal P}$ using $\widehat{\sigma}_i$
as defined in equation (\ref{Eq:sigma_Roukema_modified})
are plotted as a function of $L$.
The maximum of ${\cal P}$ is computed over all orientations within
a small $L$ interval.
}
\end{figure}

\begin{figure}
\begin{center}
\begin{minipage}{10cm}
\vspace*{-25pt}
\hspace*{-20pt}\includegraphics[width=10cm]{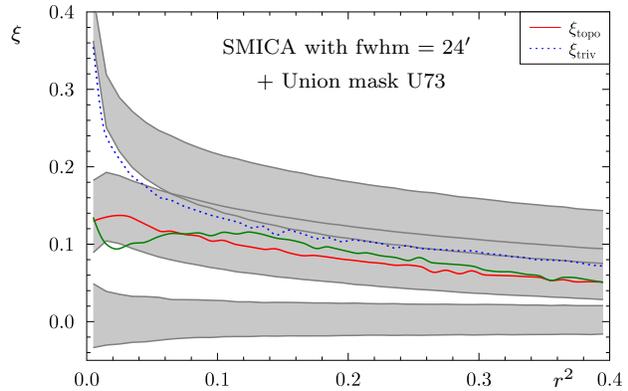}
\put(-190,145){SMICA with fwhm = $24'$}
\put(-180,132){ + Union mask U73}
\put(-65,19){$r^2$}
\put(-270,150){$\xi$}
\end{minipage}
\vspace*{-25pt}
\end{center}
\caption{\label{Fig:corr_topo_planck_256_24_cmp_2_candidates}
The correlation $\xi_{\hbox{\scriptsize topo}}(r)$ is shown for the
$L=4.41$ candidate as in figure \ref{Fig:corr_topo_planck_wmap}(a),
but, in addition, the correlation of the $L=4.52$ configuration is also
plotted which belongs to the second maximum shown in
figure \ref{Fig:P_as_function_of_L}.
The latter correlation is enhanced for $r^2\gtrsim 0.08$, but is reduced
for smaller values of $r^2$ compared to $\xi_{\hbox{\scriptsize topo}}(r)$
belonging to the $L=4.41$ candidate.
}
\end{figure}

The configurations generated by the MCMC algorithm contain not only the
three Euler angles, but also the side length $L$ of the cubic 3-torus cell.
The $L$ values are now grouped into bins and within each bin,
the maximal value of ${\cal P}$ is searched regardless of the
orientations of the cells.
The resulting curve, shown in figure \ref{Fig:P_as_function_of_L},
displays a bimodal distribution.
The largest value of ${\cal P}$ belongs to the $L=4.41$ candidate
discussed above.
The second maximum occurs at $L=4.52$ but the orientation is nearly
identical.
The fact that the orientations are very similar, can be inferred
from figure \ref{Fig:mollweide_axis}, where the ${\cal P}$ values
of all members of the MCMC chain are taken into account.
Another orientation would lead to a plot with further six directions
with high probabilities, but this is not the case.
The correlations $\xi_{\hbox{\scriptsize topo}}(r)$ belonging to the two models
with $L=4.41$ and $L=4.52$ are compared in
figure \ref{Fig:corr_topo_planck_256_24_cmp_2_candidates}.
For $r^2\gtrsim 0.08$ the topological correlation is even larger
for the $L=4.52$ model than for the $L=4.41$ configuration,
but at small values of $r^2$ the correlation is reduced.
Since the chosen pseudo-probability ${\cal P}$ tries to emphasize
the peak structure at $r=0$ in $\xi_{\hbox{\scriptsize triv}}(r)$,
as discussed at the beginning of section \ref{Sec:Comparison_with_Planck},
the model with $L=4.41$ is slightly preferred.
Although other versions of ${\cal P}$ would favour other models,
the general conclusion would be unchanged
that the orientation given in table \ref{Tab:Torus_axes} leads
to large correlations $\xi_{\hbox{\scriptsize topo}}(r)$ with $r^2\leq 0.4$
for values of $L$ in the range shown in figure \ref{Fig:P_as_function_of_L}.
Furthermore, the topological correlation $\xi_{\hbox{\scriptsize topo}}(r)$
matches the expectation derived from the 100 CMB simulations of
the cubic 3-torus topology.

\begin{figure}
\begin{center}
\begin{minipage}{10cm}
\vspace*{-25pt}
\hspace*{-20pt}\includegraphics[width=10cm]{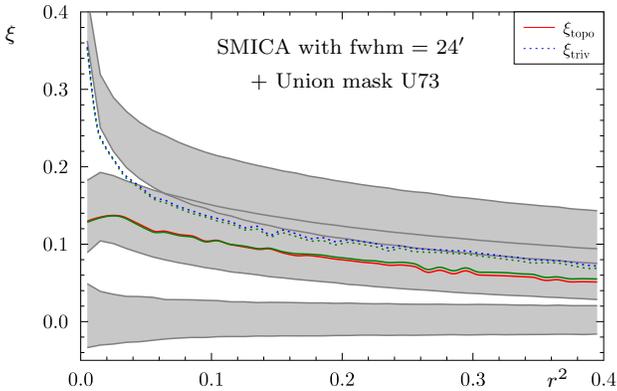}
\put(-190,145){SMICA with fwhm = $24'$}
\put(-180,132){ + Union mask U73}
\put(-65,19){$r^2$}
\put(-270,150){$\xi$}
\end{minipage}
\vspace*{-25pt}
\end{center}
\caption{\label{Fig:corr_topo_planck_256_24_2013_2015_candidates}
The correlations $\xi_{\hbox{\scriptsize topo}}(r)$ and
$\xi_{\hbox{\scriptsize triv}}(r)$ are shown for the
$L=4.41$ candidate as in figure \ref{Fig:corr_topo_planck_wmap}(a).
In addition, the correlations $\xi_{\hbox{\scriptsize topo}}(r)$
and $\xi_{\hbox{\scriptsize triv}}(r)$ are plotted that are obtained
from the Planck 2015 data using the Union mask U73.
Both correlations are almost identical to those obtained from
the 2013 data,
although the 2015 correlation $\xi_{\hbox{\scriptsize topo}}(r)$ is
slightly enhanced for $r^2\gtrsim 0.2$.
}
\end{figure}

Although the analysis of this paper is based on the Planck 2013 data,
it is interesting to see whether the recently published Planck 2015 data
\citep{Planck_2015_IX}
also display the enhanced spatial correlations for the $L=4.41$ candidate.
The temperature differences outside the masks between the 2013 and
2015 data sets are stated to be of the order of at most $\sim 10\mu K$ 
by \cite{Planck_2015_IX}.
Thus, only modest differences are expected, and this is confirmed
by figure \ref{Fig:corr_topo_planck_256_24_2013_2015_candidates},
where $\xi_{\hbox{\scriptsize topo}}(r)$ is plotted for the $L=4.41$ candidate
using the 2013 SMICA map as above, as well as using the
recently published 2015 SMICA map.
The corresponding curves of $\xi_{\hbox{\scriptsize topo}}(r)$
and $\xi_{\hbox{\scriptsize triv}}(r)$ are each nearly indistinguishable.
A MCMC chain of length 100\,000 is generated which explores the
parameter space around the $L=4.41$ candidate based on the 2015 data.
The orientation of the fundamental symmetry axes of the cubic 3-torus candidate
turns out as depicted in figure \ref{Fig:mollweide_axis} which is obtained
from the 2013 data.
The axes obtained from the 2015 data agree with those given
in table \ref{Tab:Torus_axes}.

\section{Summary and discussion}
\label{Sec:Summary}

\cite{Roukema_et_al_2008a} put forward the idea
that a spatial-correlation analysis can serve as a tool
for detecting the topology of our Universe.
The topological correlation $\xi_{\hbox{\scriptsize topo}}(r)$,
defined in equation (\ref{Eq:corr_topo}),
should agree with the correlation $\xi_{\hbox{\scriptsize triv}}(r)$,
equation (\ref{Eq:corr_triv}),
if the hypothesized topology used for the computation of
$\xi_{\hbox{\scriptsize topo}}(r)$ matches the true topology.

The spatial-correlation analysis is applied to the
cubic 3-torus topology using the Planck 2013 data.
CMB simulations for the cubic 3-torus topology are generated and
analysed with respect to the spatial correlation in
section \ref{Sec:Test_for_torus}.
This allows to estimate the level of correlations
that is to be expected if a suitable configuration is found.
It is shown that the agreement between $\xi_{\hbox{\scriptsize topo}}(r)$
and $\xi_{\hbox{\scriptsize triv}}(r)$ is very good only under very
idealized conditions
where the CMB simulations take only the usual Sachs-Wolfe contribution
into account.
Although a realistic CMB simulation based on the Boltzmann physics leads to
enhanced topological correlations $\xi_{\hbox{\scriptsize topo}}(r)$,
they nevertheless lie appreciably below $\xi_{\hbox{\scriptsize triv}}(r)$
as shown in figure \ref{Fig:corr_topo_torus_no_dipol}.

The cubic 3-torus configuration
which was found by \cite{Aurich_2008} as having a large topological
correlation $\xi_{\hbox{\scriptsize topo}}(r)$ using the WMAP five-year data,
is reanalysed in section \ref{Sec:Comparison_with_Planck}
using the Planck 2013 data.
Figure \ref{Fig:corr_topo_l_3.66} shows that the topological correlation 
is indeed of the expected magnitude for not too small distances $r$,
but possesses relatively small values of $\xi_{\hbox{\scriptsize topo}}(r)$
for small distances and does not match the increasing behaviour found in
CMB 3-torus simulations.
An improved search for configurations of the cubic 3-torus topology
found, however, a new candidate,
which has a topological correlation $\xi_{\hbox{\scriptsize topo}}(r)$
that is very close to the mean topological correlation obtained from
100 torus simulations.
This configuration is denoted as the $L=4.41$ candidate and its correlation
is shown in figures \ref{Fig:corr_topo_planck_wmap},
\ref{Fig:corr_topo_planck_512_05}, and
\ref{Fig:corr_topo_planck_256_24_2013_2015_candidates}.
The orientation of its fundamental symmetry axes is visualized in
figure \ref{Fig:mollweide_axis} and listed in table \ref{Tab:Torus_axes}.

Although the agreement with the expected correlation is amazing,
one must keep in mind that the spatial-correlation method has a
relative high false-positive rate.
To determine this rate, it is necessary to carry out the
search for a matching topology with CMB simulations generated for
the trivial topology.
This demanding computation is done by \cite{Aurich_2008}
for 10 such CMB simulations, and it turns out
that one of these 10 simulations generates a high correlation by chance.
Based on this limited number of simulations,
a false-positive rate of the order of one to ten is expected,
and \cite{Aurich_2008} concludes that $\sim 10\%\dots 20\%$ of
the simulations would produce such a false-positive detection.
It should be noted that these simulations are carried out
in the lower HEALPix resolution $N_{\hbox{\scriptsize side}}=128$
in contrast to this paper,
where the higher resolution $N_{\hbox{\scriptsize side}}=256$
or even $N_{\hbox{\scriptsize side}}=512$ is used.
Nevertheless, the order of the false-positive detection rate should
carry over to the resolution used in this paper,
since the resolution does not significantly change the
spatial correlations as the comparison of
figure \ref{Fig:corr_topo_planck_wmap}(a)
with figure \ref{Fig:corr_topo_planck_512_05} shows.
The former is computed for $N_{\hbox{\scriptsize side}}=256$ and
$\hbox{fwhm}=24'$ and the latter for $N_{\hbox{\scriptsize side}}=512$ and
$\hbox{fwhm}=7'$, and a very similar result is obtained.
In addition, figures \ref{Fig:corr_topo_l_3.66}(a) and
\ref{Fig:corr_topo_l_3.66}(b) allow a comparison between the resolutions
$\hbox{fwhm}=60'$ and $\hbox{fwhm}=24'$ again with very similar
spatial correlations.
Therefore, the false-positive rate is of the order of one to ten.

Furthermore, the Planck 2013 data are analysed with respect to
a non-trivial topology by the \cite{Planck_Topo_2013} and no
convincing signal is detected by the matched circle test
and by using off-diagonal correlations.
The application of the covariance matrix for detecting a toroidal universe
is discussed by \cite{Kunz_et_al_2006,Kunz_et_al_2007,Phillips_Kogut_2006}.
It should be noted that the use of the covariance matrix
$C_{lm}^{l'm'} := \left< a_{lm} a_{l'm'}^\star \right>$
can fail to detect a topology if the phases of $a_{lm}$ are
not accurately enough determined as
shown by \cite{Aurich_Janzer_Lustig_Steiner_2007}
where the influence of a mask is investigated.
Nevertheless, it remains the point that no hints in favour
of a toroidal universe come from the matched circle test
and from off-diagonal correlations.

This points to a false-positive detection.
However, it might be that the simulations underestimate the
deteriorating effects,
which disturb the CMB with respect to the clean signal
that would be obtained from the usual Sachs-Wolfe contribution.
A hint in this direction is the fact that the cross-correlation
of the CMB radiation with the large-scale galaxy distribution
is significantly stronger than expected
\citep{Granett_Neyrinck_Szapudi_2008,Aiola_Kosowsky_Wang_2014}.
This points to the possibility that the late-time integrated Sachs-Wolfe effect
is larger than predicted by the $\Lambda$CDM concordance model.
The CMB simulations are based on the  $\Lambda$CDM concordance model
and thus do not include such currently inexplicable contributions.
However, the $\Lambda$CDM model is used to discriminate between
a spurious signal
and a signal corresponding to a genuine detection of a non-trivial topology.
This estimation of the detection level might be flawed due to
additional correlations
that do not have their origin at the surface of last scattering.

There remains the question whether the 3-torus topology is hidden
in the CMB data.
But even if the enhanced spatial correlation of the $L=4.41$ candidate
is not due to a cubic 3-torus topology,
it is nevertheless striking that there exists an orientation of a
torus cell, for which a large correlation in the CMB data is found.


\section*{Acknowledgements}

HEALPix [healpix.jpl.nasa.gov]
\citep{Gorski_Hivon_Banday_Wandelt_Hansen_Reinecke_Bartelmann_2005} and
the Planck as well as WMAP data from the LAMBDA website (lambda.gsfc.nasa.gov)
were used in this work.



\bibliography{../bib_astro}

\begin{thebibliography}{}

\bibitem[{Aiola} et~al., 2015]{Aiola_Kosowsky_Wang_2014}
{Aiola}, S., {Kosowsky}, A., and {Wang}, B. (2015).
\newblock {Gaussian Approximation of Peak Values in the Integrated Sachs-Wolfe
  Effect}.
\newblock {\em \prd}, 91(4):043510, arXiv:1410.6138 [astro-ph.CO].

\bibitem[{Atrio-Barandela} et~al., 2014]{Atrio_Barandela_et_al_2014}
{Atrio-Barandela}, F., {Kashlinsky}, A., {Ebeling}, H., {Fixsen}, D.~J., and
  {Kocevski}, D. (2014).
\newblock {Probing the Dark Flow signal in WMAP 9 yr and PLANCK cosmic
  microwave background maps}.
\newblock {\em ArXiv e-prints}, arXiv:1411.4180 [astro-ph.CO].

\bibitem[{Aurich}, 2008]{Aurich_2008}
{Aurich}, R. (2008).
\newblock A spatial correlation analysis for a toroidal universe.
\newblock {\em \cqg}, 25:225017, arXiv:0803.2130 [astro-ph].

\bibitem[{Aurich} et~al., 2008]{Aurich_Janzer_Lustig_Steiner_2007}
{Aurich}, R., {Janzer}, H.~S., {Lustig}, S., and {Steiner}, F. (2008).
\newblock {Do we Live in a ''Small Universe''?}
\newblock {\em \cqg}, 25:125006, arXiv:0708.1420 [astro-ph].

\bibitem[{Aurich} et~al., 2005a]{Aurich_Lustig_Steiner_2004c}
{Aurich}, R., {Lustig}, S., and {Steiner}, F. (2005a).
\newblock {CMB} anisotropy of the {P}oincar\'e {D}odecahedron.
\newblock {\em \cqg}, 22:2061--2083, arXiv:astro-ph/0412569.

\bibitem[{Aurich} et~al., 2005b]{Aurich_Lustig_Steiner_Then_2004b}
{Aurich}, R., {Lustig}, S., {Steiner}, F., and {Then}, H. (2005b).
\newblock Indications about the shape of the {U}niverse from the {W}ilkinson
  {M}icrowave {A}nisotropy {P}robe data.
\newblock {\em \prl}, 94:021301--1--4, arXiv:astro-ph/0412407.

\bibitem[{Bennett} et~al., 2013]{Bennett_et_al_2012}
{Bennett}, C.~L., {Larson}, D., {Weiland}, J.~L., {Jarosik}, N., {Hinshaw}, G.,
  {Odegard}, N., {Smith}, K.~M., {Hill}, R.~S., {Gold}, B., {Halpern}, M.,
  {Komatsu}, E., {Nolta}, M.~R., {Page}, L., {Spergel}, D.~N., {Wollack}, E.,
  {Dunkley}, J., {Kogut}, A., {Limon}, M., {Meyer}, S.~S., {Tucker}, G.~S., and
  {Wright}, E.~L. (2013).
\newblock {Nine-Year Wilkinson Microwave Anisotropy Probe (WMAP) Observations:
  Final Maps and Results}.
\newblock {\em \apjs}, 208:20, arXiv:1212.5225 [astro-ph.CO].

\bibitem[{Copi} et~al., 2009]{Copi_Huterer_Schwarz_Starkman_2008}
{Copi}, C.~J., {Huterer}, D., {Schwarz}, D.~J., and {Starkman}, G.~D. (2009).
\newblock No large-angle correlations on the non-{G}alactic microwave sky.
\newblock {\em \mnras}, 399:295--303, arXiv:0808.3767 [astro-ph].

\bibitem[{Copi} et~al., 2015]{Copi_Huterer_Schwarz_Starkman_2013a}
{Copi}, C.~J., {Huterer}, D., {Schwarz}, D.~J., and {Starkman}, G.~D. (2015).
\newblock {Lack of large-angle TT correlations persists in WMAP and Planck}.
\newblock {\em \mnras}, 451:2978--2985, arXiv:1310.3831 [astro-ph.CO].

\bibitem[{Cornish} et~al., 1998]{Cornish_Spergel_Starkman_1998b}
{Cornish}, N.~J., {Spergel}, D.~N., and {Starkman}, G.~D. (1998).
\newblock Circles in the sky: finding topology with the microwave background
  radiation.
\newblock {\em \cqg}, 15:2657--2670, arXiv:astro-ph/9801212.

\bibitem[{Fujii} and {Yoshii}, 2011]{Fujii_Yoshii_2011}
{Fujii}, H. and {Yoshii}, Y. (2011).
\newblock {An improved cosmic crystallography method to detect holonomies in
  flat spaces}.
\newblock {\em \aap}, 529:A121, arXiv:1103.1466 [astro-ph.CO].

\bibitem[{Gold} et~al., 2009]{Gold_et_al_2008}
{Gold}, B., {Bennett}, C.~L., {Hill}, R.~S., {Hinshaw}, G., {Odegard}, N.,
  {Page}, L., {Spergel}, D.~N., {Weiland}, J.~L., {Dunkley}, J., {Halpern}, M.,
  {Jarosik}, N., {Kogut}, A., {Komatsu}, E., {Larson}, D., {Meyer}, S.~S.,
  {Nolta}, M.~R., {Wollack}, E., and {Wright}, E.~L. (2009).
\newblock {Five-Year Wilkinson Microwave Anisotropy Probe (WMAP) Observations:
  Galactic Foreground Emission}.
\newblock {\em \apjs}, 180:265--282, arXiv:0803.0715 [astro-ph].

\bibitem[{Gomero} and {Rebou{\c c}as}, 2003]{Gomero_Reboucas_2003}
{Gomero}, G.~I. and {Rebou{\c c}as}, M.~J. (2003).
\newblock {Detectability of cosmic topology in flat universes}.
\newblock {\em Physics Letters A}, 311:319--330, arXiv:gr-qc/0202094.

\bibitem[{G\'orski} et~al.,
  2005]{Gorski_Hivon_Banday_Wandelt_Hansen_Reinecke_Bartelmann_2005}
{G\'orski}, K.~M., {Hivon}, E., {Banday}, A.~J., {Wandelt}, B.~D., {Hansen},
  F.~K., {Reinecke}, M., and {Bartelmann}, M. (2005).
\newblock {HEALPix: A Framework for High-Resolution Discretization and Fast
  Analysis of Data Distributed on the Sphere}.
\newblock {\em \apj}, 622:759--771.
\newblock HEALPix web-site: http://healpix.jpl.nasa.gov/.

\bibitem[{Granett} et~al., 2008]{Granett_Neyrinck_Szapudi_2008}
{Granett}, B.~R., {Neyrinck}, M.~C., and {Szapudi}, I. (2008).
\newblock {An Imprint of Superstructures on the Microwave Background due to the
  Integrated Sachs-Wolfe Effect}.
\newblock {\em \apjl}, 683:L99--L102, arXiv:0805.3695 [astro-ph].

\bibitem[{Hinshaw} et~al., 1996]{Hinshaw_et_al_1996}
{Hinshaw}, G., {Banday}, A.~J., {Bennett}, C.~L., {G\'orski}, K.~M., {Kogut},
  A., {Lineweaver}, C.~H., {Smoot}, G.~F., and {Wright}, E.~L. (1996).
\newblock {Two-Point Correlations in the COBE DMR Four-Year Anisotropy Maps}.
\newblock {\em \apjl}, 464:L25--L28.

\bibitem[{Kim} and {Naselsky}, 2011]{Kim_Naselsky_2011}
{Kim}, J. and {Naselsky}, P. (2011).
\newblock {Lack of Angular Correlation and Odd-parity Preference in Cosmic
  Microwave Background Data}.
\newblock {\em \apj}, 739:79, arXiv:1011.0377 [astro-ph.CO].

\bibitem[{Kunz} et~al., 2006]{Kunz_et_al_2006}
{Kunz}, M., {Aghanim}, N., {Cayon}, L., {Forni}, O., {Riazuelo}, A., and
  {Uzan}, J.~P. (2006).
\newblock Constraining topology in harmonic space.
\newblock {\em \prd}, 73(2):023511--+, arXiv:astro-ph/0510164.

\bibitem[{Kunz} et~al., 2008]{Kunz_et_al_2007}
{Kunz}, M., {Aghanim}, N., {Riazuelo}, A., and {Forni}, O. (2008).
\newblock On the detectability of non-trivial topologies.
\newblock {\em \prd}, 77:023525, arXiv:astro-ph/0704.3076.

\bibitem[{Lachi\`eze-Rey} and {Luminet}, 1995]{Lachieze-Rey_Luminet_1995}
{Lachi\`eze-Rey}, M. and {Luminet}, J.-P. (1995).
\newblock Cosmic topology.
\newblock {\em Physics Report}, 254:135--214.

\bibitem[{Levin}, 2002]{Levin_2002}
{Levin}, J. (2002).
\newblock Topology and the cosmic microwave background.
\newblock {\em Physics Report}, 365:251--333.

\bibitem[{Luminet} and {Roukema}, 1999]{Luminet_Roukema_1999}
{Luminet}, J.-P. and {Roukema}, B.~F. (1999).
\newblock {Topology of the Universe: Theory and Observation}.
\newblock In {\em NATO ASIC Proc. 541: Theoretical and Observational
  Cosmology}, page 117. Kluwer, astro-ph/9901364.

\bibitem[{Mota} et~al., 2010]{Mota_Reboucas_Tavakol_2010}
{Mota}, B., {Rebou{\c c}as}, M.~J., and {Tavakol}, R. (2010).
\newblock {Circles-in-the-sky searches and observable cosmic topology in a flat
  universe}.
\newblock {\em \prd}, 81:103516, arXiv:1002.0834 [astro-ph.CO].

\bibitem[{Mota} et~al., 2011]{Mota_Reboucas_Tavakol_2011}
{Mota}, B., {Rebou{\c c}as}, M.~J., and {Tavakol}, R. (2011).
\newblock {What can the detection of a single pair of circles-in-the-sky tell
  us about the geometry and topology of the Universe?}
\newblock {\em \prd}, 84:083507, arXiv:1108.2842 [astro-ph.CO].

\bibitem[{Phillips} and {Kogut}, 2006]{Phillips_Kogut_2006}
{Phillips}, N.~G. and {Kogut}, A. (2006).
\newblock {Constraints on the Topology of the Universe from the Wilkinson
  Microwave Anisotropy Probe First-Year Sky Maps}.
\newblock {\em \apj}, 645:820--825, arXiv:astro-ph/0404400.

\bibitem[{Planck Collaboration} et~al., 2015]{Planck_2015_IX}
{Planck Collaboration}, {Adam}, R., {Ade}, P.~A.~R., {Aghanim}, N., {Arnaud},
  M., {Ashdown}, M., {Aumont}, J., {Baccigalupi}, C., {Banday}, A.~J.,
  {Barreiro}, R.~B., and et~al. (2015).
\newblock {Planck 2015 results. IX. Diffuse component separation: CMB maps}.
\newblock {\em ArXiv e-prints}, 1502.05956.

\bibitem[{Planck Collaboration} et~al., 2014a]{Planck_Overview_2013}
{Planck Collaboration}, {Ade}, P.~A.~R., {Aghanim}, N., {Alves}, M.~I.~R.,
  {Armitage-Caplan}, C., {Arnaud}, M., {Ashdown}, M., {Atrio-Barandela}, F.,
  {Aumont}, J., {Aussel}, H., and et~al. (2014a).
\newblock {Planck 2013 results. I. Overview of products and scientific
  results}.
\newblock {\em \aap}, 571:A1, arXiv:1303.5062.

\bibitem[{Planck Collaboration} et~al., 2014b]{Planck_Isotropy_2013}
{Planck Collaboration}, {Ade}, P.~A.~R., {Aghanim}, N., {Armitage-Caplan}, C.,
  {Arnaud}, M., {Ashdown}, M., {Atrio-Barandela}, F., {Aumont}, J.,
  {Baccigalupi}, C., {Banday}, A.~J., and et~al. (2014b).
\newblock {Planck 2013 results. XXIII. Isotropy and statistics of the CMB}.
\newblock {\em \aap}, 571:A23, arXiv:1303.5083 [astro-ph.CO].

\bibitem[{Planck Collaboration} et~al., 2014c]{Planck_Topo_2013}
{Planck Collaboration}, {Ade}, P.~A.~R., {Aghanim}, N., {Armitage-Caplan}, C.,
  {Arnaud}, M., {Ashdown}, M., {Atrio-Barandela}, F., {Aumont}, J.,
  {Baccigalupi}, C., {Banday}, A.~J., and et~al. (2014c).
\newblock {Planck 2013 results. XXVI. Background geometry and topology of the
  Universe}.
\newblock {\em \aap}, 571:A26, arXiv:1303.5086 [astro-ph.CO].

\bibitem[{Rebou\c{c}as} and {Gomero}, 2004]{Reboucas_Gomero_2004}
{Rebou\c{c}as}, M.~J. and {Gomero}, G.~I. (2004).
\newblock {Cosmic Topology: a Brief Overview}.
\newblock {\em Braz.~J.~Phys.}, 34:1358--1366, arXiv:astro-ph/0402324.

\bibitem[{Roukema} et~al., 2008]{Roukema_et_al_2008a}
{Roukema}, B.~F., {Buli\'nski}, Z., {Szaniewska}, A., and {Gaudin}, N.~E.
  (2008).
\newblock {The optimal phase of the generalised Poincare Dodecahedral Space
  hypothesis implied by the spatial cross-correlation function of the WMAP sky
  maps}.
\newblock {\em \aap}, 486:55, arXiv:0801.0006 [astro-ph].

\bibitem[{Roukema} et~al., 2014]{Roukema_France_Kazimierczak_Buchert_2013}
{Roukema}, B.~F., {France}, M.~J., {Kazimierczak}, T.~A., and {Buchert}, T.
  (2014).
\newblock {Deep redshift topological lensing: strategies for the $T^3$
  candidate}.
\newblock {\em \mnras}, 437:1096--1108, arXiv:1302.4425 [astro-ph.CO].

\end{thebibliography}
\bibliographystyle{hapalike}


\end{document}